\documentclass[sigconf]{acmart}

\copyrightyear{2023}
\acmYear{2023}
\setcopyright{rightsretained}
\setcopyright{rightsretained}
\acmConference[KDD '23]{Proceedings of the 29th ACM SIGKDD Conference on Knowledge Discovery and Data Mining}{August 6--10, 2023}{Long Beach, CA, USA}
\acmBooktitle{Proceedings of the 29th ACM SIGKDD Conference on Knowledge Discovery and Data Mining (KDD '23), August 6--10, 2023, Long Beach, CA, USA}
\acmDOI{10.1145/3580305.3599249}
\acmISBN{979-8-4007-0103-0/23/08}

\usepackage{balance}

\usepackage{microtype}
\usepackage{graphicx}
\usepackage{subfigure}
\usepackage{booktabs} %
\usepackage{hyperref}

\usepackage{amsmath}
\usepackage{mathtools}
\usepackage{amsthm}

\usepackage[capitalize,noabbrev]{cleveref}

\usepackage{enumitem}
\usepackage{xspace}
\usepackage{autobreak}
\usepackage{multicol,multirow}
\usepackage{colortbl}
\usepackage{bm}
\usepackage{algorithm}
\usepackage{algorithmic}

\newenvironment{itemize*}%
  {\begin{itemize}%
    \setlength{\itemsep}{2pt}%
    \setlength{\parskip}{2pt}}%
  {\end{itemize}}
\newenvironment{enumerate*}%
  {\begin{enumerate}%
    \setlength{\itemsep}{2pt}%
    \setlength{\parskip}{2pt}}%
  {\end{enumerate}}
\newenvironment{enumerate**}%
  {\begin{enumerate}%
    \setlength{\itemsep}{0pt}%
    \setlength{\parskip}{0pt}}%
  {\end{enumerate}}

\theoremstyle{plain}

\theoremstyle{definition}

\theoremstyle{remark}

\newcommand{\method}{\texttt{LSSAMP}\xspace}

\makeatletter
\gdef\@copyrightpermission{
 \begin{minipage}{0.3\columnwidth}
  \href{https://creativecommons.org/licenses/by/4.0/}{\includegraphics[width=0.90\textwidth]{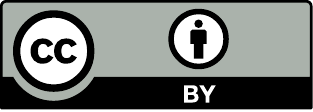}}
 \end{minipage}\hfill
 \begin{minipage}{0.7\columnwidth}
  \href{https://creativecommons.org/licenses/by/4.0/}{This work is licensed under a Creative Commons Attribution International 4.0 License.}
 \end{minipage}
 \vspace{5pt}
}
\makeatother

\begin{document}

\title{Accelerating Antimicrobial Peptide Discovery with Latent Structure}

\author{Danqing Wang}
\authornote{Part of this work was done when the author was in ByteDance Research.}
\affiliation{%
  \institution{University of California, Santa Barbara}
  \state{California}
  \country{USA}
}
\email{danqingwang@ucsb.edu}

\author{Zeyu Wen}
\authornotemark[1]
\affiliation{
 \institution{Huazhong University of Science and Technology}
 \city{Wuhan}
 \country{China}
}
\email{D201780029@hust.edu.cn}

\author{Fei Ye}
\affiliation{%
 \institution{ByteDance Research}
 \city{Shanghai}
 \country{China}
 }
 \email{yefei.joyce@bytedance.com}

\author{Lei Li}
\authornotemark[1]
\authornote{Contributed equally. Corresponding to LL and HZ.}
\affiliation{%
  \institution{University of California, Santa Barbara}
  \state{California}
  \country{USA}
}
\email{leili@cs.ucsb.edu}

\author{Hao Zhou}
\authornotemark[1]
\authornotemark[2]
\affiliation{%
 \institution{Institute for AI Industry Research, Tsinghua University}
 \city{Beijing}
 \country{China}
 }
\email{zhouhao@air.tsinghua.edu.cn}

\begin{abstract}
Antimicrobial peptides (AMPs) are promising therapeutic approaches against drug-resistant pathogens.
Recently, deep generative models are used to discover new AMPs.
However, previous studies mainly focus on peptide sequence attributes and do not consider crucial structure information.
In this paper, we propose a latent sequence-structure model for designing AMPs (\method).
\method exploits multi-scale vector quantization in the latent space to represent secondary structures (e.g. alpha helix and beta sheet). 
By sampling in the latent space, \method can simultaneously generate peptides with ideal sequence attributes and secondary structures.
Experimental results show that the peptides generated by \method have a high probability of antimicrobial activity. Our wet laboratory experiments verified that two of the 21 candidates exhibit strong antimicrobial activity. The code is released at https://github.com/dqwang122/LSSAMP.

\end{abstract}



\maketitle

\section{Introduction}
\label{sec:intro}
In recent years, the development of neural networks for drug discovery has attracted increasing attention. It can facilitate the discovery of potential therapies and reduce the time and cost of drug development~\citep{stokes2020deep}. Great success has been achieved in applying deep generative models to accelerate the discovery of potential drug-like molecules~\citep{jin2018learning,shi2019graphaf,schwalbe2020generative,xie2020mars}. 

Antimicrobial peptides (AMPs) are one of the most promising emerging therapeutic agents to replace antibiotics. They are short proteins that can kill bacteria by destroying the bacterial membrane~\citep{aronica2021computational,cardoso2020computer}. Compared with the chemical interactions between antibiotics and bacteria that can be avoided by bacterial evolution, this physical mechanism is more difficult to resist.

\begin{figure}
    \centering
    \includegraphics[width=1\linewidth]{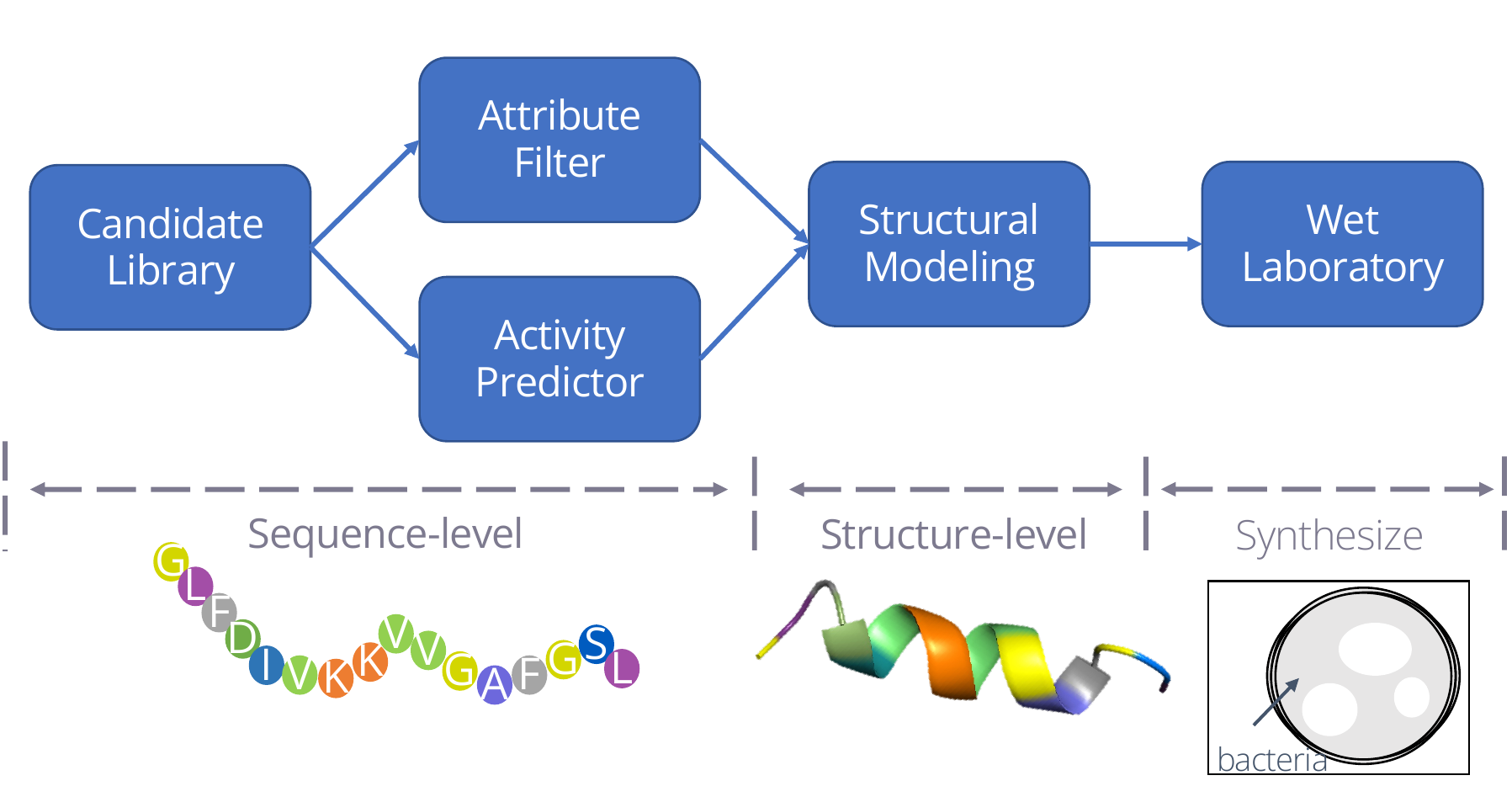}
    \caption{The overview of AMP discovery. The first two steps focus on sequence attributes and the third models the structure. The final step is to verify the antimicrobial activity by inhibiting the growth of bacteria. The grep region is the bacterial suspension and the white one means the bacteria concentration in this region is small.}
    \label{fig:procedure}
\end{figure}

A typical antimicrobial discovery process usually consists of four steps, as shown in Figure \ref{fig:procedure}. First, a candidate library is built based on the existing AMPs database. These candidates can be created by applying manual heuristic approaches or training deep generative models. Then, several sequence-based filters are created to screen candidate peptides based on different chemical features, including computational metrics and predictive models trained to estimate ideal properties. After that, to ensure that these sequences can fold into appropriate structures with biological functions, the structure of these sequences will be modeled using peptide structure predictors such as PEPFold 3~\citep{shen2014improved} and molecule dynamics simulations will be performed. Finally, the filtered sequences will be synthesized and tested in the wet laboratory.

Recently, deep generative models have achieved great success in accelerating AMP discovery. They use sequence attributes to control the generation and directly generate peptides with ideal attributes directly~\citep{das2018pepcvae,das2021accelerated,van2021ampgan}. However, these studies only consider sequence features and ignore the structure-activity relationship. The generated sequences still need to be fed into structure predictors and checked manually, which slows down the discovery process. Besides, the structure also plays an important role in determining biological attributes, which in turn facilitates attribute control~\citep{chen2019simulation,torres2018structure,tucker2018discovery}.

In this paper, we incorporate the structure information into the generative model and propose a Latent Sequence-Structure model for AntiMicrobial Peptide (\method). It maps the sequence features and secondary structures into the same latent space and samples the peptides with ideal sequence compositions and structures.\method controls the generation in a more fine-grained manner by assigning a latent variable to each position instead of a continuous variable to control the attributes of the whole sequence. We employ a multi-scale vector quantized-variational autoencoder (VQ-VAE)~\citep{van2017neural} to capture patterns of different lengths on sequences and structures. During the generation process, \method samples from the latent space and generates a peptide sequence with its secondary structure. The experimental results via public AMP predictors show that the peptides generated by \method have a high probability of AMP. We further conduct a comprehensive qualitative analysis, which indicates that our model captures the sequence and structure distribution. We select 21 generated peptides and conduct wet laboratory experiments, and find that 2 of them have high antimicrobial activity against Gram-negative bacteria.

To conclude, our contributions are as follows:

\begin{itemize}
  \item We propose \method, a sequence-structure generative model that combines secondary structure information into the generation, which can further accelerate AMP discovery.
  \item We develop a multi-scale VQ-VAE to control the generation in a fine-grained manner and map patterns in sequences and structures into the same latent space.
  \item Experimental results of AMP predictors show that \method generates peptides with high probabilities of AMP. Moreover, 2 of 21 generated peptides show strong antimicrobial activities in wet laboratory experiments.
\end{itemize}

\section{Related work}
\label{sec:related}
\textbf{Antimicrobial Peptide Generation}
Traditional methods for AMP discovery can be divided into three approaches~\citep{torres2019toward}: 
(i) \textit{Pattern recognition algorithms} first builds an antimicrobial sequential pattern database from existing AMPs. Each time a template peptide is chosen to substitute local fragments with those patterns~\citep{loose2006linguistic,porto2018joker}. 
(ii) \textit{Genetic algorithms} use the AMP database to design some antimicrobial activity functions and optimize ancestral sequences with these functions ~\citep{maccari2013antimicrobial}. 
(iii) \textit{Molecular modeling and molecular dynamics methods} build 3D models of peptides and evaluate their antimicrobial activity by the interaction between peptides and the bacterial membrane~\citep{matyus2007computer,bolintineanu2011computational}. 
Pattern recognition and genetic algorithm bottleneck on representing patterns, and the modeling and dynamics method is computationally expensive and time-consuming. 

Deep generative models take a rapid growth in recent years. 
\citet{dean2020variational} encodes the peptide into the latent space and interpolates across a predictive vector between a known AMP and its scrambled version to generate novel peptides.
The PepCVAE~\citep{das2018pepcvae} and CLaSS~\citep{das2021accelerated} employ the variational auto-encoder model to generate sequences. 
The AMPGAN~\citep{van2021ampgan} uses the generative adversarial network to generate new peptide sequences with a discriminator distinguishing the real AMPs and artificial ones. 
To our knowledge, this is the first study to incorporate secondary structure information into the generative phase, which is conducive to efficiently generating well-structured sequences with desired properties.

\textbf{Sequence Generation via VQ-VAE}
The variational auto-encoders (VAEs) were first proposed by~\citet{kingma2013auto} for image generation, and then widely applied to sequence generation tasks such as language modeling~\citep{bowman2016generating}, paraphrase generation~\citep{gupta2018deep}, machine translation~\citep{bao2019generating} and so on.
Instead of mapping the input to a continuous latent space in VAE, the vector quantized-variational autoencoder (VQ-VAE)~\citep{van2017neural} learns the codebook to obtain a discrete latent representation. It can avoid issues of posterior collapse while has comparable performance with VAEs. Based on it, \citet{razavi2019generating} uses a multi-scale hierarchical organization to capture global and local features for image generation. \citet{bao2021non} learns implicit categorical information of target words with VQ-VAE and models the categorical sequence with conditional random fields in non-autoregressive machine translation. In this paper, we employ the multi-scale vector quantized technique to obtain the discrete representation for each position of the peptide.

\section{Latent Sequence-Structure Model}
\label{sec:method}
In this section, we first introduce the background of AMP discovery and discuss the limitation of existing generative models. Then we introduce the Latent Sequence-Structure model for AMP (\method), which uses the multi-scale VQ-VAE to co-design sequence and secondary structure within the same latent space.

\begin{table*}[tb]
\setlength{\tabcolsep}{3pt}
  \centering
    \begin{tabular}{lccccc}
    \toprule
      & \multicolumn{1}{c}{\textbf{Uniq}} & \multicolumn{1}{c}{\textbf{C}} & \multicolumn{1}{c}{\textbf{H }} & \multicolumn{1}{c}{\textbf{uH}} & \multicolumn{1}{c}{\textbf{Combination}} \\
    \midrule
    \textbf{VAE}~\citep{dean2020variational} & 475 & 18.45\% $\pm$ 2.92\% & 2.68\% $\pm$ 3.28\% & -2.78\% $\pm$ 1.64\% & 0.29\% $\pm$ 0.74\% \\
    \textbf{AMP-GAN}~\citep{van2021ampgan} & 1966 & 2.79\% $\pm$ 0.50\% & 2.16\% $\pm$ 0.34\% & -2.29\% $\pm$ 0.53\% & 0.17\% $\pm$ 0.35\% \\
    \textbf{PepCVAE}~\citep{das2018pepcvae} & 208 & 3.87\% $\pm$ 1.58\% & -1.93\% $\pm$ 1.61\% & 1.01\% $\pm$ 2.80\% & 3.93\% $\pm$ 1.82\% \\
    \textbf{MLPeptide}~\citep{capecchi2021machine} & 2106 & -2.48\% $\pm$ 0.39\% & 2.01\% $\pm$ 0.57\% & 9.24\% $\pm$ 1.22\% & 1.12\% $\pm$ 0.38\% \\
    \bottomrule
    \end{tabular}%
  \caption{The delta ratio of the sequence properties filtered by the secondary structures. \textbf{Uniq} is the uniq peptide number in 5000 generated sequences. \textbf{C}, \textbf{H}, \textbf{uH} correspond to charge, hydrophobicity, hydrophobic moment. \textbf{Combination} is the percentage of satisfying three valid ranges for AMPs at the same time.}
  \label{tab:filtered}%
\end{table*}%

\subsection{Background}
As shown in Figure \ref{fig:procedure}, a typical AMP discovery includes the sequence-level attribute and the structure-level modeling before the wet laboratory experiments. Existing deep generative models have shown promise in accelerating AMP discovery by considering the sequence-level attributes during the generation.
However, they still need to check and filter the structures with external tools after the sequence generation, which makes the process less efficient. For example, \citet{van2021ampgan} manually check the generated peptides and only got 12 AMP candidates with the ideal cationic and helical structure. \citet{capecchi2021machine} required an extra secondary structure predictor (SPIDER3~\cite{heffernan2018single}) to filter the generated peptides based on the percentage of the predicted a-helix structure fraction.

Moreover, there is a close relationship between the structure and activity of peptides. We investigate the effect of secondary structure on sequence properties by filtering the generated sequences based on the proportion of $\alpha$-helices, which is the most common secondary structure in AMPs. In Table \ref{tab:filtered}, we use three sequence attributes (\textit{charge}, \textit{hydrophobicity}, \textit{hydrophobic moment}) that are crucial for AMP mechanism to evaluate generation performance~\citep{yeaman2003mechanisms,gidalevitz2003interaction,wimley2010describing}\footnote{The definition of these three attributes can be found in Section \ref{sec:attr}.}. 

The ratio in Table \ref{tab:filtered} is the difference in performance before and after the secondary structure filter. We can find that most of the results are improved by limiting alpha-helical structures. 
The results show that by controlling the structure, the sequence properties can be improved. Thus, incorporating the structure information into generative models can not only accelerate discovery by combining all the steps before the wet laboratory but also improve the sequence properties and make the generative process more efficient. 

To address these challenges, we combine the secondary structure with sequence attributes in AMP discovery and generate peptides with ideal sequence attributes and secondary structures simultaneously.

\paragraph{Notation}
A peptide\footnote{Here, we use the peptide to refer to the oligopeptide ($<$ 20 amino acids) and the polypeptide ($<$ 50 amino acids).} with length $L$ can be denoted by $\mathbf{x} =\{x_{1}, x_{2}, \cdots, x_{L}\}$ and $x_{i} \in \mathcal{V}_a$ belongs to one of the 20 common amino acids, which is also called a \textit{residue}.
The \textit{secondary structure} is used to describe the local form of the 3D structure of the peptide. It can be annotated as $\mathbf{y}=\{y_{1}, y_{2}, \cdots, y_{L}\}$, where $y_{i} \in \mathcal{V}_s$ is from 8 secondary structure types~\footnote{The three alpha helices are denoted as H, G, and I based on their angles. The two beta sheets are divided into E and T by shape. The others are random coil structures~\citep{kabsch1983dictionary}.}. The goal is to generate peptide candidates with high antimicrobial activities to accelerate the AMP discovery process.

\begin{figure}
    \centering
    \includegraphics[width=1\linewidth]{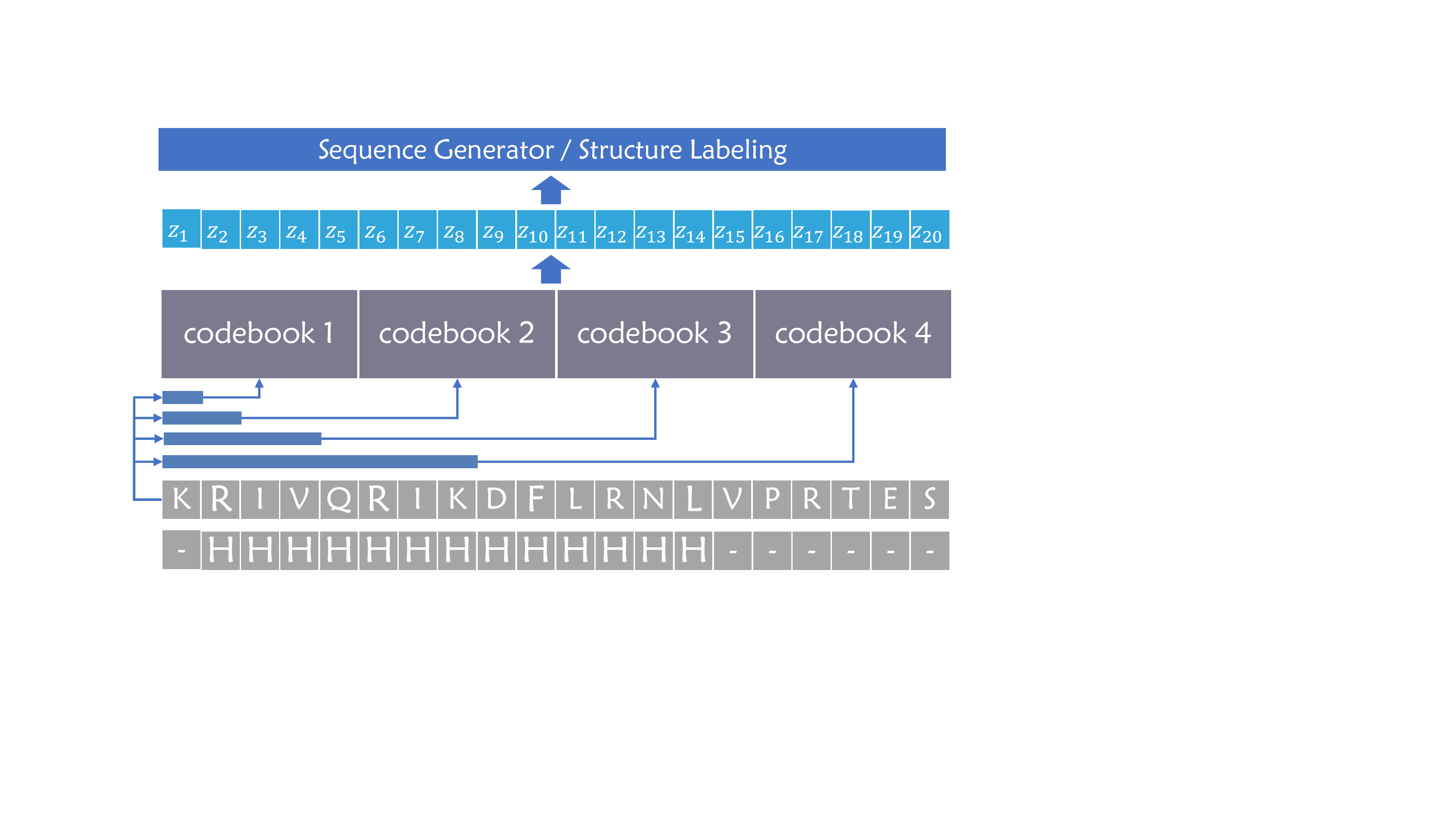}
    \caption{The encoder of \method. Here, we use $N=4$ pattern selectors to select local patterns with different scales and use the corresponding codebooks to obtain discrete latent variables $z_{i}$ for each position. The number of selectors is further discussed in Section \ref{sec:book}.}
    \label{fig:model}
\end{figure}

\subsection{VAE-based AMP Models}
\label{sec:vae}

Given a sequence $\mathbf{x}$, the variational auto-encoders assume that it depends on a continuous latent variable $\mathbf{z}$. Thus the likelihood can be denoted by:
\begin{align}
p(\mathbf{x}) = \int p(\mathbf{z})p(\mathbf{x}|\mathbf{z})d\mathbf{z}.
\end{align}
The controlled sequence generation incorporates the attribute $a$ and models the conditional probability $p(\mathbf{x}|a)$. Previous work such as PepCVAE~\citep{das2018pepcvae} assumes that $\mathbf{z}$ and $a$ are independent and $p(\mathbf{x}|a)=\int_{\mathbf{z}} p(\mathbf{z})p(\mathbf{x}|\mathbf{z}, a)d\mathbf{z}$, while CLaSS~\citep{das2021accelerated} models the dependency between $\mathbf{z}$ and $a$ by $p(\mathbf{x}|a)=\int_{\mathbf{z}} p(\mathbf{z}|a)p(\mathbf{x}|\mathbf{z})d\mathbf{z}$.

The vanilla VAE are usually trained in an auto-encoder framework with regularization. The encoder parameterizes an approximate posterior distribution $q_{\phi}(\mathbf{z}|\mathbf{x})$ and the decoder $p_\theta(\mathbf{x}|\mathbf{z})$ reconstructs $\mathbf{x}$ based on the latent $\mathbf{z}$. The models optimizes a evidence lower bound ($\text{ELBO}$):
\begin{align}
\label{eqn:elbo}
ELBO = E_{q_{\phi}(\mathbf{z}|\mathbf{x})}[log(p_\theta(\mathbf{x}|\mathbf{z}))] - \text{KL}(q_{\phi}(\mathbf{z}|\mathbf{x}) || p(\mathbf{z})),
\end{align}
where the $E_{q_{\phi}(\mathbf{z}|\mathbf{x})}[log(p_\theta(\mathbf{x}|\mathbf{z}))]$ is the reconstruction loss and the $\text{KL}$ divergence is the regularization. For the conditional generation, the attributes are directly fed to the decoder with latent variable $\mathbf{z}$ for $p_{\theta}(\mathbf{x}|\mathbf{z}, a)$, or trained on the latent space to get an attribute-conditioned posterior distribution $p(\mathbf{z}|a)$. The VAE-based peptide generative models are first trained on the unsupervised peptide or protein sequences and then finetuned with a few sequences with biological attribute labels.

\subsection{\method}
\label{sec:lssamp}
To capture the sequence and the secondary structure feature simultaneously, we would like to model the joint distribution $p(\mathbf{x}, \mathbf{y})$.
Based on Eqn \ref{eqn:elbo}, the training objective is:
\begin{align}
ELBO = E_{q_{\phi}(\mathbf{z}|\mathbf{x}, \mathbf{y})}[log(p_\theta(\mathbf{x},\mathbf{y}|\mathbf{z}))] - \text{KL}(q_{\phi}(\mathbf{z}|\mathbf{x}, \mathbf{y}) || p(\mathbf{z})).
\end{align}
For sequence $\mathbf{x}$ and secondary structure $\mathbf{y}$, we assume that they are independent given the latent variable $\mathbf{z}$. Thus the joint distribution can be written as $p(\mathbf{x}, \mathbf{y}) = \int_z p(\mathbf{x}|\mathbf{z}) p(\mathbf{y}|\mathbf{z}) p(\mathbf{z}) d\mathbf{z}$. 
Moreover, since a peptide has a deterministic secondary structure given its amino acid sequence, we further approximate $q_{\phi}(\mathbf{z}|\mathbf{x}, \mathbf{y})$ with $q_{\Phi}(\mathbf{z}|\mathbf{x})$. Therefore, the training objective is written as:
\begin{align}
\label{eqn:elbo_joint}
ELBO = E_{q_{\Phi}(\mathbf{z}|\mathbf{x})}[log(p_{\theta_a}(\mathbf{x}|\mathbf{z})) + log(p_{\theta_s}(\mathbf{y}|\mathbf{z}))]  \nonumber \\
- \text{KL}(q_{\Phi}(\mathbf{z}|\mathbf{x}) || p(\mathbf{z})).
\end{align}

For fine-grained control over each position, we assign one latent variable $z_{i}$ for each $x_{i}$ instead of a continuous $\mathbf{z}$ for the whole sequence. 
Since it is computationally intractable to sum continuous latent variables over the sequence, we use VQ-VAE~\citep{van2017neural} to lookup the discrete embedding vector $\mathbf{z}_{q} = \{z_{q}(x_{1}), \cdots, z_{q}(x_{L})\}$ for each position by vector quantization.

Specifically, the original latent variable $z_{e}(x_{i}) \in \mathbb{R}^d$ will be replaced by the codebook entry $z_{q}(x_{i}) \in \mathbb{R}^d$ via a nearest neighbors lookup from the codebook $\mathbf{B} \in \mathbb{R}^{K \times d}$ :
\begin{align}
\label{eqn:lookup}
    & z_{q}(x_{i})  = e_{k}, \text{and } k  = \operatorname{argmin}_{j \in \{1,\cdots,K\}}  \left\|{z_{e}(x_{i})-e_{j}}\right\|_{2}.
\end{align}
Here, $K$ is the slot number of the codebook and $d$ is the dimension of the codebook entry $e$. Then, the decoder will take $z_{q}(x_{i})$ as its input. So $\text{KL}(q_{\Phi}(\mathbf{z}|\mathbf{x}) || p(\mathbf{z}))$ in Eqn \ref{eqn:elbo_joint} is replaced by:
\begin{align}
    \left\|\operatorname{sg}\left[z_{e}(x_{i})\right]-z_{q}(x_{i})\right\|_{2}^{2}
    +\beta\left\|z_{e}(x_{i})-\operatorname{sg}[z_{q}(x_{i})]\right\|_{2}^{2},
\end{align}
measuring the difference between the  original latent variable $z_{e}(x_{i})$ and the nearest codebook entry, rather than the KL divergence between two continuous distributions. Here, $sg(\cdot)$ is the stop gradient operator, which becomes 0 at the backward pass. $\beta$ is the commit coefficient to control the codebook loss. 

For the sequence feature, we use the reconstruction loss to learn $\theta_a$, and for the secondary structure, we view it as an 8-category labeling task. Therefore, the first term in Eqn \ref{eqn:elbo_joint} becomes:
\begin{align}
    \sum_{i=1}^{L} \log p_{\theta_a}\left(x_{i} | z_{q}(x_{i})\right) +
    \sum_{i=1}^{L} \log p_{\theta_s}\left(y_{i} | z_{q}(x_{i})\right)
\end{align}

However, the structure motifs are often longer than chain patterns. Therefore, we establish multiple codebooks to capture features of various scales.

\textbf{Multi-scale VQ-VAE} 
The structure motifs are often longer than sequence patterns. For example, a valid $\alpha$- helix contains at least 4 residues and may be longer than 12. However, sequence patterns with specific biological functions are much shorter, usually between 1 and 8 residues. In order to capture these features and map them into the same latent space, we first apply $N$ multi-scale pattern selectors $F_{n}$ on $z_e$ and get $z_{e_{n}}$. Then, we establish multiple codebooks $B_n \in \mathbb{R}^{K \times d}$ and use Eqn. \ref{eqn:lookup} to look up the nearest codebook embedding $z_{q_{n}}(x_{i})$. 
We share the codebooks between sequence reconstruction and secondary structure prediction to capture common features and relationships between the residue and its structure. 
The concatenated multi-scale codebook embedding is fed to the sequence generator:
\begin{align}
    & z_{q}(x_{i}) = \|_{n \in N}{z_{q_{n}}(x_{i})},
\end{align}

The total training objective is composed of the reconstruction loss, the labeling loss, and the codebook loss, which can be denoted by:
\begin{align}
\label{eqn:loss}
    ELMO = \sum_{i=1}^{L} \log p_{\theta_a}\left(x_{i} | z_{q}(x_{i})\right) +
    \sum_{i=1}^{L} \log p_{\theta_s}\left(y_{i} | z_{q}(x_{i})\right) \nonumber \\
    - \sum_{n=1}^{N}\left[\left\|\operatorname{sg}\left[z_{e_{n}}(x_{i})\right]-z_{q_{n}}(x_{i})\right\|_{2}^{2}
    +\beta\left\|z_{e_{n}}(x_{i})-\operatorname{sg}[z_{q_{n}}(x_{i})]\right\|_{2}^{2}\right]
\end{align}
The framework is shown in Figure \ref{fig:model}.

\begin{algorithm}[t]
\footnotesize
\caption{Training and Sampling phase of \method}
\label{alg:algorthm}
\begin{algorithmic}[1]
\REQUIRE 
A protein dataset $D_{r}$, a peptide dataset with secondary structure $D_{s}$, and the AMP dataset $D_{amp}$. The model $M$=\{$\theta_{a}$, $\theta_{s}$, $B_{n}$ \} with $N$ codebooks. A set of $N$ prior models $M_{prior_{n}}$.
\STATE Train on $D_{r}$ to optimize $\theta_{a}$ and $B_{n}$.
\STATE Train on $D_{s}$ to optimize $\theta_{a}$, $\theta_{s}$, $B_{n}$ via Eqn. \ref{eqn:loss}.
\STATE Finetune $M$ on $D_{amp}$ via Eqn. \ref{eqn:loss}.

\FOR{ each codebook $n = 1,2,\cdots,N$ }
    \STATE Create an empty dataset $C_{n}$.
    \FOR{$x_{i} \in D_{amp}$}
        \STATE Get $n$-th codebook index $k_{i_{n}}$ of $x_{i}$ via Eqn. \ref{eqn:lookup} and save it to $C_{n}$
    \ENDFOR
    \STATE Train an auto-regressive language model $M_{\text{prior}_{n}}$ on $C_{n}$.
\ENDFOR

\end{algorithmic}
\end{algorithm}

\textbf{Training} 
\label{sec:train}
As VAE-based generative models, we first train \method in an unsupervised manner with protein sequences, which is similar to the original ELBO in Eqn \ref{eqn:elbo}. Then, we incorporate the structure information by jointly training on a smaller protein dataset with secondary structure annotation. Finally, we finetune our model on the AMP dataset to capture the specific AMP characteristics. The whole training process is described in Algorithm \ref{alg:algorthm}.

Following \citet{kaiser2018fast}, we use Exponential Moving Average (EMA) to update the embedding vectors in the codebooks. Specifically, we keep a count $c_k$ measuring the number of times that the embedding vector $e_{k}$ is chosen as the nearest neighbor of $z_{e}(x_{i})$ via Eqn. \ref{eqn:lookup}. Thus, the counts are updated with a sort of momentum: $c_{k} \gets \lambda c_{k} + (1 - \lambda) \sum_{i} \mathbb{I}[z_{q}(x_{i})=e_{k}]$, with the embedding $e_{k}$ being updated as $e_{k} \gets \lambda e_{k} + (1 - \lambda) \sum_{i} \frac{\mathbb{I}[z_{q}(x_{i})=e_{k}]z_{e}(x_{i})}{c_{k}}$. Here, $\lambda$ is the decay parameter.

\textbf{Modeling dependency between position} To model the dependency between $z_{1:L}$, we build an auto-regressive model on the index sequence of each codebook:
\begin{align}
    p(\mathbf{z}) = p(z_{1:L}) = \prod_{i=1}^{L} p(z_i | z_{<i}).
\end{align}
Specifically, we train Transformer-based language models $M_{\text{prior}_{n}}$ based on the index sequences from Eqn. \ref{eqn:lookup} for each codebook $n$ .

\textbf{Sampling} We sample several index sequences from the prior models for each codebook $n$, and then look up the codebook to get the embedding vector $\mathbf{z}_{q_{n}}$. Finally, $\mathbf{z}_{q_{n}}$ is fed to the decoder to generate the sequence with its secondary structure. We also further control the secondary structure by heuristic structure patterns to improve the generation quality.

\section{Experiment}
\label{sec:exper}
We first describe our experiment settings (Section \ref{sec:setup}) and introduce the automatic evaluation metrics (Section \ref{sec:metric}) and results (Section \ref{sec:res}). Then, we verify the antimicrobial activity in the wet laboratory (Section \ref{sec:wet}). Last but not least, we conduct an in-depth analysis (Section \ref{sec:analysis}) to understand \method and discuss its limitation (Section \ref{sec:limitation}).  

\begin{table*}[ht]\setlength{\tabcolsep}{4pt}
  \centering
    \begin{tabular}{lcccccccc}
    \toprule
      & \multicolumn{1}{c}{\textbf{SVM}} & \multicolumn{1}{c}{\textbf{RF}} & \multicolumn{1}{c}{\textbf{DA}} & \multicolumn{1}{c}{\textbf{Scanner}} & \multicolumn{1}{c}{\textbf{AMPMIC}} & \multicolumn{1}{c}{\textbf{IAMPE}} & \multicolumn{1}{c}{\textbf{amPEP}} & \multicolumn{1}{c}{\textbf{Average}} \\
    \midrule
    \textbf{APD} & 87.78\% & 91.24\% & 86.24\% & 94.66\% & 98.42\% & 97.83\% & 91.50\% & 92.52\% \\
    \textbf{Decoy} & 17.43\% & 13.71\% & 16.04\% & 0.25\% & 18.07\% & 23.53\% & 52.92\% & 20.28\% \\
    \midrule
    \textbf{Random $p=0.1$} & 86.06\% & 86.12\% & 84.01\% & 93.23\% & 79.14\% & 95.60\% & 91.74\% & 87.99\% \\
    \textbf{Random $p=0.2$} & 76.66\% & 76.64\% & 74.83\% & 86.95\% & 68.57\% & 91.14\% & 87.89\% & 80.38\% \\
    \textbf{VAE}~\citep{dean2020variational} & 24.90\% & 15.30\% & 13.83\% & 15.12\% & 15.25\% & 40.31\% & 24.30\% & 21.29\% \\
    \textbf{AMP-GAN}~\citep{van2021ampgan} & 78.62\% & 87.29\% & 83.82\% & 82.17\% & 89.58\% & 93.88\% & 80.52\% & 85.13\% \\
    \textbf{PepCVAE}~\citep{das2018pepcvae} & 82.84\% & 85.96\% & 93.33\% & 85.44\% & \textbf{98.44}\% & \textbf{98.14}\% & 80.77\% & 89.27\% \\
    \textbf{MLPeptide}~\citep{capecchi2021machine} & 90.43\% & 92.55\% & 93.08\% & \textbf{93.72}\% & 96.34\% & 97.05\% & 91.37\% & 93.51\% \\
    \midrule
    \textbf{\method} & \textbf{92.03}\% & \textbf{92.60}\% & \textbf{93.45}\% & 91.52\% & 95.84\% & 96.64\% & \textbf{93.23}\% & \textbf{93.62}\% \\
    \textbf{\method w/o cond} & 78.98\% & 80.24\% & 80.01\% & 86.73\% & 83.81\% & 93.80\% & 85.32\% & 84.13\% \\
    \bottomrule
    \end{tabular}%
  \caption{The percentage of generated sequences being predicted as AMP. The classifiers are described in Section \ref{sec:cls}. The first part is the prediction results on AMP and non-AMP dataset as the reference. The bold ones are the best model results.}
  \label{tab:cls}%
\end{table*}%

\subsection{Experiment Setup}
\label{sec:setup}

\textbf{Dataset}
The Universal Protein Resource (UniProt)\footnote{https://www.uniprot.org/} is a comprehensive protein dataset. We download reviewed protein sequences (550k) with the limitation of 100 in length as $D_{r}$ (57k examples). Then we use a community reimplementation of AlphaFold\citep{alquraishi2019alphafold}, which is called ProSPr\footnote{https://github.com/dellacortelab/prospr/tree/prospr1}~\citep{billings2019prospr} to predict the secondary structure for $D_{r}$. After filtering some low-quality examples, we obtain $D_{s}$ with 46k examples, including both sequence and secondary structure information. For the antimicrobial peptide dataset, we download from Antimicrobial Peptide Database (\textbf{APD})\footnote{https://aps.unmc.edu/}~\citep{wang2016apd3} and filter repeated ones to get 3222 AMPs as $D_{amp}$.
We randomly extract 3,000 examples as validation and 3,000 as the test on $D_{r}$
and $D_{s}$. For $D_{amp}$, the size of validation and test is both 100.
Following~\citet{veltri2018deep}, we create a decoy set of negative examples without antimicrobial activities for comparison. It removes peptide sequences with antimicrobial activity from Uniprot, and sequences with length $< 10$ or $> 40$, resulting in 2021 non-AMP sequences (\textbf{Decoy}).

\textbf{Baseline}
Traditional methods usually randomly replace several residues on existing AMPs and conduct biological experiments on them. Thus, we use \textbf{Random} baseline to represent the method of replacing residue randomly with a probability $p$. 
Following \citet{dean2020variational}, we use \textbf{VAE} to embed the peptides into the latent space and sample latent variable $z$ from the standard Gaussian distribution $p \sim N(0,1)$. For a fair comparison, we use the same Transformer architecture as our model \method and train on the Uniprot $D_{r}$ and APD dataset $D_{amp}$. 
\textbf{AMP-GAN} is proposed by \citet{van2021ampgan}, which uses a BiCGAN architecture with convolution layers. It consists of three parts: the generator, discriminator, and encoder. The generator and discriminator share the same encoder. It is trained on 49k false negative sequences from UniProt and 7k positive AMP sequences.
\textbf{PepCVAE} is a semi-VAE generative model that concatenates the attribute features to the latent variable for conditional generation~\citep{das2018pepcvae}. Since the authors did not release their code, we use the model architecture from \citet{hu2017toward} and modify the reproduced code\footnote{https://github.com/wiseodd/controlled-text-generation} for AMPs, as described in their paper. The original paper uses 93k sequences from UniProt and 7960/6948 positive/negative AMPs for training. For comparison, we use UniProt dataset $D_{r}$ and ADP dataset $D_{amp}$ to train it.
\textbf{MLPeptide}~\citep{capecchi2021machine} is RNN-based generator. It is first trained on 3580 AMPs and then transferred to specific bacteria.
\textbf{\method} is implemented as described in Section \ref{sec:lssamp} and the detailed hyperparameters are attached in Appendix \ref{sec:implementation}.

\subsection{Automatic Evaluation Metric}
\label{sec:metric}

Following previous work~\citep{das2020accelerating,van2021ampgan}, we use open-source AMP prediction tools to estimate the AMP probability of the generated sequence. Since these open-source AMP predictors are trained and report results in different AMP datasets, we use them to predict sequences in APD and decoy datasets as a reference of their performance. We also evaluate the generative diversity of these models.

\begin{table*}[ht]\setlength{\tabcolsep}{3pt}
  \centering
    \begin{tabular}{lccccc}
    \toprule
      & \multicolumn{1}{c}{\textbf{Uniq}} & \multicolumn{1}{c}{\textbf{C}} & \multicolumn{1}{c}{\textbf{H }} & \multicolumn{1}{c}{\textbf{uH}} & \multicolumn{1}{c}{\textbf{Combination}} \\
    \midrule
    \textbf{APD} & 3222 & 68.75\% & 27.96\% & 4.72\% & 6.15\% \\
    \textbf{Decoy} & 2020 & 21.83\% & 8.81\% & 1.98\% & 0.10\% \\
    \midrule
    \textbf{Random $p = 0.1$} & 4978 & 65.86\% $\pm$ 0.19\% & \textbf{26.80}\% $\pm$ 0.23\% & 23.10\% $\pm$ 0.58\% & 4.38\% $\pm$ 0.16\% \\
    \textbf{Random $p = 0.2$} & 5000 & 62.13\% $\pm$ 0.39\% & 24.87\% $\pm$ 0.29\% & 20.79\% $\pm$ 0.76\% & 2.47\% $\pm$ 0.17\% \\
    \textbf{VAE} & 4988 & 38.00\% $\pm$ 0.36\% & 21.07\% $\pm$ 0.58\% & 12.43\% $\pm$ 0.66\% & 0.34\% $\pm$ 0.11\% \\
    \textbf{AMP-GAN} & 4976 & \textbf{87.66}\% $\pm$ 0.45\% & 17.31\% $\pm$ 0.74\% & 23.45\% $\pm$ 0.73\% & 1.92\% $\pm$ 0.05\% \\
    \textbf{PepCVAE} & 1346 & 15.61\% $\pm$ 0.06\% & 14.54\% $\pm$ 0.55\% & 11.65\% $\pm$ 0.23\% & 2.75\% $\pm$ 0.25\% \\
    \textbf{MLPeptide} & 4486 & 77.95\% $\pm$ 0.72\% & 8.11\% $\pm$ 0.27\% & 32.91\% $\pm$ 0.60\% & 2.90\% $\pm$ 0.16\% \\
    \midrule
    \textbf{\method} & 4876 & 81.88\% $\pm$ 0.31\% & 25.06\% $\pm$ 0.45\% & \textbf{37.10}\% $\pm$ 0.33\% & \textbf{6.26}\% $\pm$ 0.07\% \\
    \textbf{\method w/o cond} & 4903 & 82.04\% $\pm$ 0.42\% & 21.32\% $\pm$ 0.34\% & 30.51\% $\pm$ 0.51\% & 4.46\% $\pm$ 0.20\% \\
    \bottomrule
    \end{tabular}%
  \caption{Sequence attributes of generated sequences. We use the percentage of peptides meeting the valid range (Appendix \ref{sec:apd}) to measure the performance. \textbf{Uniq} is the number of unique generated sequences. \textbf{C}, \textbf{H}, \textbf{uH} correspond to charge, hydrophobicity, hydrophobic moment. The best results are bold.}
  \label{tab:results}%
\end{table*}%

\subsubsection{AMP Classifiers}
\label{sec:cls}
\citet{thomas2010camp} trained on the AMP database of 3782 sequences with random forest (\textbf{RF}), discriminant analysis (\textbf{DA}), support vector machines (\textbf{SVM})\footnote{http://www.camp3.bicnirrh.res.in/prediction.php}, and artificial neural network (ANN)\footnote{We drop the ANN model because its accuracy on APD is low (82.83\%).} respectively.
AMP Scanner v2\footnote{https://www.dveltri.com/ascan/v2/ascan.html}~\citep{veltri2018deep}, short as \textbf{Scanner},  is a CNN-\&LSTM-based deep neural network trained on 1778 AMPs picked from APD.
\textbf{AMPMIC}\footnote{https://github.com/zswitten/Antimicrobial-Peptides}~\citep{witten2019deep} trained a CNN-based regression model on 6760 unique sequences and 51345 MIC measurement to predict MIC values.
\textbf{IAMPE}\footnote{http://cbb1.ut.ac.ir/AMPClassifier/Index}~\citep{kavousi2020iampe} is a model based on Xtreme Gradient Boosting. It achieves the highest correct prediction rate on a set of ten more recent AMPs~\citep{aronica2021computational}.
\textbf{ampPEP}\footnote{https://github.com/tlawrence3/amPEPpy}~\citep{lawrence2021ampeppy} is a random forest based model which is trained on 3268 AMPs. It has the best performance across multiple datasets~\citep{aronica2021computational}.

\subsubsection{Sequence Attributes} 
\label{sec:attr}
Following the previous AMP design~\citep{das2018pepcvae,van2021ampgan,capecchi2021machine,das2021accelerated}, we use the three sequence attributes to evaluate the generation performance: Charge(C), Hydrophobicity(H), Hydrophobic Momentum (uH). Here, we use $x_i$ to denote the $i$-th residue and $S$ to indicate the sequence.

\textbf{Charge} is important because the bacterial membrane usually takes the negative charge and peptides with the positive charge are more likely to bind with the membrane. We only take integer charges into consideration. The whole charge of the peptide sequence $S$ is defined as the sum of the charge of all its residues $C(x_{i})$ at pH 7.4, which is $C(S)=\sum_{x_{i} \in S}{C(x_{i})}$.

\textbf{Hydrophobicity} reflects the tendency to bind lipids on the bacterial membrane. A peptide with a high hydrophobicity is easy to move from the solution environment to the bacterial membrane.We use the hydrophobicity scale $H(x_{i})$ in \citet{eisenberg1984hydrophobic} to calculate the hydrophobicity of a sequence, which is $H(S) = \sum_{x_{i} \in S}{H(x_{i})}$.

\textbf{Hydrophobic Momentum}. It is viewed as the measure of amphipathicity, indicating the ability of the peptide to bind water and lipid simultaneously. It is a definitive feature of antimicrobial peptides\citep{hancock2002role}. Hydrophobic momentum $uH(S, \theta)$ is defined by \citet{eisenberg1984hydrophobic}. The hydrophobic momentum is determined by the hydrophobicity $H(x_{i})$ of each residue $x_{i}$, along with the angle $\theta$ between residues. The angle can be estimated by the secondary structure. 
$\theta$ is $100^{\circ}$ for the $\alpha$-helix structure, and $180^{\circ}$ for $\beta$-sheet.
\begin{align}
uH(S, \theta) = \sqrt{R_{cos}^2(S, \theta) + R_{sin}^2(S, \theta)}, \\
R_{cos}(S, \theta) = \sum_{x_{i} \in S}{H(x_{i}) * cos(i* \theta)}, \\
R_{sin}(S, \theta) = \sum_{x_{i} \in S}{H(x_{i}) * sin(i* \theta)}.
\end{align}

For each peptide, we calculate the above attributes to measure its antimicrobial activity. For comparison, we draw the distribution on the APD and decoy dataset and select a range for each attribute based on the biological mechanism (Appendix \ref{sec:apd}). 
We use the percentage of peptides in each attribute range to exploit the generation performance and use \textbf{Combination} to measure the percentage of peptides that satisfy three conditions at the same time.

\subsubsection{Novelty}
\label{sec:novelty}
To measure the novelty of the generated peptides, we define three evaluation metrics: Uniqueness, Diversity, and Similarity. \textbf{Uniqueness} is the percentage of unique peptides in the generation phase. \textbf{Diversity} measures the similarity among the generated peptides. We calculate the Levenshtein distance~\citep{levenshtein1966binary} between every two sequences and normalize it by the sequence length. Then we average the normalized distance to get the mean as its diversity. The higher the diversity, the more dissimilar the generated peptides are. \textbf{Novelty} is the difference between the generated peptides and the training AMP set. For each generated sequence, we search the training set for a peptide that has the smallest Levenshtein distance from it and normalizes the distance according to its length. We calculate the average length as the Novelty.

\begin{table}[tb]
\setlength{\tabcolsep}{1.5pt}
  \centering
    \begin{tabular}{lccc}
    \toprule
      & \multicolumn{1}{c}{\textbf{Uniqueness $\uparrow$}} & \multicolumn{1}{c}{\textbf{Diversity $\uparrow$}} & \multicolumn{1}{c}{\textbf{Novelty $\uparrow$}} \\
    \midrule
    \textbf{Random $p=0.1$} & 0.995 $\pm$ 0.000 & 0.871 $\pm$ 0.021 & 0.078 $\pm$ 0.001  \\
    \textbf{Random $p=0.2$} & \textbf{0.999} $\pm$ 0.000 & 0.971 $\pm$ 0.022 & 0.160 $\pm$ 0.001 \\
    \textbf{VAE} & 0.986  $\pm$ 0.001 & \textbf{1.011} $\pm$ 0.038 & \textbf{0.584} $\pm$ 0.002 \\
    \textbf{AMP-GAN} & 0.995 $\pm$ 0.001 & 0.907 $\pm$ 0.023 & 0.565 $\pm$ 0.007 \\
    \textbf{PepCVAE} & 0.265 $\pm$ 0.006 & 0.367 $\pm$ 0.007 & 0.423 $\pm$ 0.005 \\
    \textbf{MLPeptide} & 0.900 $\pm$ 0.003 & 0.850 $\pm$ 0.016 & 0.416 $\pm$ 0.010 \\
    \midrule
    \textbf{\method} & 0.981 $\pm$ 0.001 & 0.878 $\pm$ 0.018 & 0.503 $\pm$ 0.005 \\
    \textbf{\method w/o cond} & 0.976 $\pm$ 0.002 & 0.901 $\pm$ 0.013 & 0.515 $\pm$ 0.008 \\
    \bottomrule
    \end{tabular}%
  \caption{The novelty of the sampling. $\uparrow$ means higher is better.}
  \label{tab:novelty_full}%
\end{table}%

\begin{table}[tb]
    \centering\small
    \setlength{\tabcolsep}{3pt}
    \begin{tabular}{lcccc}
    \toprule
      & \multicolumn{1}{c}{\textbf{PPL $\downarrow$}} & \multicolumn{1}{c}{\textbf{Loss $\downarrow$}} & \multicolumn{1}{c}{\textbf{AA Acc.$\uparrow$}} & \multicolumn{1}{c}{\textbf{SS Acc.$\uparrow$}} \\
    \midrule
    \textbf{\method} & \textbf{3.24} $\pm$ 0.16 & \textbf{1.17} $\pm$ 0.05 & 99.79 $\pm$ 0.20 & \textbf{87.20} $\pm$ 0.62 \\
    w/o $D_{r}$ & 11.56 $\pm$ 3.81 & 2.45 $\pm$ 0.94 & 66.06$\pm$ 0.67 & 82.78 $\pm$ 0.57 \\
    w/o $D_{s}$ & 3.83 $\pm$ 0.32 & 1.34 $\pm$ 0.04 & 99.58$\pm$ 0.26 & 85.87$\pm$ 0.35 \\
    w/o subbook & 3.49 $\pm$ 0.20 & 1.25 $\pm$ 0.05 & 99.86 $\pm$ 0.36 & 86.61 $\pm$ 0.95\\
    \bottomrule
    \end{tabular}%
  \caption{Ablation Study on validation set of $D_{amp}$. `w/o' indicates that we remove the module from \method. $\uparrow$ means higher is better, and $\downarrow$ is the opposite.}
  \label{tab:ablation}
\end{table}

\begin{table*}[htbp]
  \centering\setlength{\tabcolsep}{2pt}
    \begin{tabular}{clccccc}
    \toprule
    \multicolumn{1}{c}{\multirow{2}[0]{*}{No}} & \multicolumn{1}{c}{\multirow{2}[0]{*}{Sequence}} & \multicolumn{3}{c}{Activity (ug/mL) $\downarrow$} & \multirow{2}[0]{*}{Sequence identity $\downarrow$ } & \multirow{2}[0]{*}{Hemolysis/Toxicity $\downarrow$} \\
      &   & A. Baumannii & P. aeruginose & E. coli  &   &  \\
    \midrule
    P1 & GAFGNFLKNVAKKAGIYLLSIAQCKLFGTP & 16-32 & / & 32-64 & 83.30\% & Low \\
    P2 & FIGFLFKLAKKIIPSLFQTKTE & 8 & 32 & / & 75.00\% & Low \\
    \bottomrule
    \end{tabular}%
  \caption{Wet laboratory experiment results. The activity is measured by MIC and $\downarrow$ means the lower the better. `Sequence identity' measures the similarity with existing AMPs and `Hemolysis/Toxicity' measures the damage to other cells.}
  \label{tab:wet}%
\end{table*}%

\subsection{Experimental Results}
\label{sec:res}
We generate 5000 sequences for each baseline. During the generation process, we add structural restrictions on positions based on the antimicrobial mechanism. Specifically, we reject peptides with more than 30\% coil structure (`-'), which can hardly fold in the solution environment, and insert the bacterial membrane in silico screening. Besides, we limit the minimum length of a continuous helix (`H') to 4 according to physical rules. We name our model with structural control as \textbf{\method} and the model without extra conditions as \textbf{\method w/o cond}.

\textbf{AMP Prediction} The results of prediction tools are shown in Table \ref{tab:cls}. \method performs best in four of seven and has the highest average score across all classifiers, indicating its advantage over baselines. PepCVAE performs best on the AMPMIC and IAMPE predictors, however, it performs poorly on the other predictors and gets a low average score. MLPeptide performs relatively evenly across predictors, outperforming other models on only Scanner and slightly underperforming our model on the average score.
The comparison of \method and \method w/o cond indicates that adding fine-grained control on the secondary structure can further improve the generation performance. It further indicates the importance of taking the secondary structure into consideration during the AMP generation.

\textbf{Sequence Attributes}
As listed in Table \ref{tab:results}, \method outperforms all baselines on the combination percentage, which indicates that our model can generate sequences satisfying multiple properties at the same time. Besides, the combination percentage is similar to APD, which means that our model generates sequences that have a similar distribution to APD. \method tends to generate peptides with higher hydrophobicity, while AMP-GAN and MLPeptide sample more cationic sequences. It is because they were only trained on sequences and more focused on the direct amino acid attributes.
Compared with them, \method can better capture the amphiphilic indicated by the highest uH for the incorporation of structure labels.
PepCVAE inefficiently generates redundant sequences, which results in a significant decrease in the number of unique sequences. Since the percentage is calculated based on the whole generation size (5000), it leads to low performance on all attributes.
Furthermore, we can find that by further controlling the secondary structure, H, uH and Combination can be improved. This verifies that secondary structure will also affect sequence attributes.

\textbf{Novelty}
From Table \ref{tab:novelty_full}, we can see that VAE has the highest diversity and novelty. However, from Table \ref{tab:cls} and Table\ref{tab:results}, we can find that the peptides generated by VAE do not have a high probability of AMP or the ideal sequence attributes. It means that the vanilla VAE trained on AMP datasets without attribute control can hardly capture the antimicrobial features. At the same time, \method has a significant advantage over the above strong baseline PepCVAE and MLPeptide. It means that our model can generate promising AMPs with relatively high novelty. Besides, the limitation of the secondary structure will lead to a decline in diversity. However, it does not result in more redundant peptides because the uniqueness does not decrease. It indicates that the restrictions make the model capture similar local patterns, but not generate the exact same sequence.

\begin{table}[tb]
    \centering
    \setlength{\tabcolsep}{1pt}
    \begin{tabular}{lcccc}
    \toprule
      \multicolumn{1}{c}{\textbf{Codebook}} &
       \multicolumn{1}{c}{\textbf{PPL $\downarrow$}} & \multicolumn{1}{c}{\textbf{Loss $\downarrow$}} & \multicolumn{1}{c}{\textbf{AA Acc.$\uparrow$}} & \multicolumn{1}{c}{\textbf{SS Acc.$\uparrow$}} \\
    \midrule
    $[1]$ & 19.04 $\pm$ 2.84 & 2.94 $\pm$ 0.14 & 65.49 $\pm$ 3.49 & 83.41 $\pm$ 2.34 \\
    $[1,2]$ & 3.84 $\pm$ 0.09 & 1.35 $\pm$ 0.02 & 99.40 $\pm$ 0.45 & 85.39 $\pm$ 0.26 \\
    $[1,2,4]$ & 3.32 $\pm$ 0.03 & 1.20 $\pm$ 0.01 & \textbf{100.00} $\pm$ 0.00 & 85.95 $\pm$ 0.42 \\
    $[1,2,4,8]$ & \textbf{3.24} $\pm$ 0.16 & \textbf{1.17} $\pm$ 0.05 & 99.79 $\pm$ 0.20 & \textbf{87.20} $\pm$ 0.62 \\
    \bottomrule
    \end{tabular}%
    \caption{The influence of the number of codebooks on validation set of $D_{amp}$. `[1,2,4,8]' indicates that we use 4 codebooks with window sizes of 1,2,4,8. The meanings of symbols are the same as Table \ref{tab:ablation}. }
    \label{tab:codebook_full}%
\end{table}

\textbf{Ablation Study} 
We conduct the ablation study for our \method and show the results in Table \ref{tab:ablation}. \textbf{PPL} is the perplexity of generated sequences that can measure fluency. \textbf{Loss} is the model loss on the validation set. \textbf{AA Acc.} is the reconstruction accuracy of residue and \textbf{SS Acc.} is the prediction accuracy of the secondary structure. We can find that without the first training phase on $D_{r}$, the model can hardly generate valid sequences. The second phase to train the model on the large-scale secondary structure dataset $D_{s}$ will affect the prediction performance on the target AMP dataset. If we remove multiple sub-codebooks and use a single large codebook with the same size, the performance will also decline.

\subsection{Wet Laboratory Experiments}
\label{sec:wet}
We synthesized and experimentally characterize peptides designed with \method. First, we filtered the 5000 generated peptide sequences based on their physical attributes (as outlined in Section \ref{sec:attr} and Appendix \ref{sec:apd}) and employed off-the-shelf AMP classifiers to select the ones with high antimicrobial scores (as detailed in Section \ref{sec:cls}). Second, we rank the sequences according to their novelty (as described in Section \ref{sec:novelty}) and select ones with edit distance greater than 5 residues from the existing training sequences. Finally, we obtained 21 peptides and synthesized them for wet-lab experiments.

Following the previous AMP design~\citep{capecchi2021machine,das2021accelerated}, we use minimum inhibitory concentration (MIC) to indicate peptide activity, which is defined as the lowest concentration of an antibiotic that prevents the visible growth of bacteria. A lower MIC means a higher antimicrobial activity. To determine MIC, the broth microdilution method was used. A colony of bacteria was grown in LB (Lysogeny broth) medium overnight at 37 degrees under PH=7. A peptide concentration range of 0.25 to 128 mg/liter was used for MIC assay. The concentration of bacteria was quantified by measuring the absorbance at 600 nm and diluted to OD600 = 0.022 in MH medium. The sample solutions(150uL) were mixed with a 4uL diluted bacterial suspension and finally inoculated with about 5 * 10E5 CFU. The Plates were incubated at 37 degrees until satisfactory growth ~18h. For each test, two columns of plates were reserved for sterile control (broth only) and growth control (broth with bacterial inoculum, no antibiotics). The MIC was defined as the lowest concentration of the peptide dendrimer that inhibited the growth of bacteria visible after treatment with MTT.

We tested their antimicrobial activities against three panels of Gram-negative bacteria (\textit{A. Baumannii}, \textit{P. aeruginose}, \textit{E. coli}), which cost about 30 days. 
As shown in Table \ref{tab:wet}, two peptides were both found to be effective against A. Baumannii. \textbf{P2} against P. aeruginose and \textbf{P1} against for E.coli also showed activity. Besides, these two newly discovered AMPs differ from existing AMPs and had low toxicity, which means they are promising new therapeutic agents.
The wet-lab experiment results demonstrate \method can effectively find AMP candidates and reduce the time.

\begin{table*}[htbp]\setlength{\tabcolsep}{4pt}
  \centering
    \begin{tabular}{lllccc}
    \toprule
    \textbf{ID} & \multicolumn{1}{c}{\textbf{Sequence}} & \multicolumn{1}{c}{\textbf{Secondary Structure}} & \multicolumn{1}{c}{\textbf{C}} & \multicolumn{1}{c}{\textbf{H}} & \multicolumn{1}{c}{\textbf{uH}} \\
    \midrule
    Y1 & FLPLVRVWAKLI &  --HHHHHHHHHH & 2.0 & 0.471 & 0.723 \\
    Y2 & FLSTVPYVAFKVVPTLFCPIAKTC &  --HHHHHHHHHHHHHHHHHHHT-- & 2.0 & 0.446 & 1.812 \\
    Y3 & FFGVLARGIKSVVKHVMGLLMG &  --HHHHHHHHHHHHHHHHHH-- & 3.0 & 0.420 & 0.549 \\
    Y4 & GVLPAFKQYLPGIMKIIVKF &  --HHHHHHHHHHHHHHH--- & 3.0 & 0.419 & 0.523 \\
    Y5 & VFTLLGAIIHHLGNFVKRFSHVF &  -HHHHHHHHHHHHHHHHHHHH-- & 2.0 & 0.416 & 0.514 \\
    Y6 & FVPGLIKAAVGIGYTIFCKISKACYQ &  --HHHHHHHHHHHHHHHHHHHT---- & 3.0 & 0.394 & 1.815 \\
    Y7 & ALWCQMLTGIGKLAGKA &  --HHHHHHHHHHHHHHH & 2.0 & 0.344 & 0.506 \\
    Y8 & LLTRIIVGAISAVTSLIKKS &  --HHHHHHHHHHHHHHHH-- & 3.0 & 0.334 & 0.531 \\
    Y9 & FLSVIKGVWAASLPKQFCAVTAKC &  --HHHHHHHHHHHHHHHHHHHT-- & 3.0 & 0.334 & 0.660 \\
    Y10 & FLNPIIKIATQILVTAIKCFLKKC &  --HHHHHHHHHHHHHHHHHHHT-- & 4.0 & 0.334 & 1.940 \\
    \bottomrule
    \end{tabular}%
  \caption{10 generated peptides. `H' is the alpha-helix, `T' is the Turn and `-' is the coil.}
  \label{tab:case}%
\end{table*}%

\begin{figure}[ht]
  \centering
  \subfigure[Y1]{
    \label{fig:seq1}
    \includegraphics[width=0.20\linewidth]{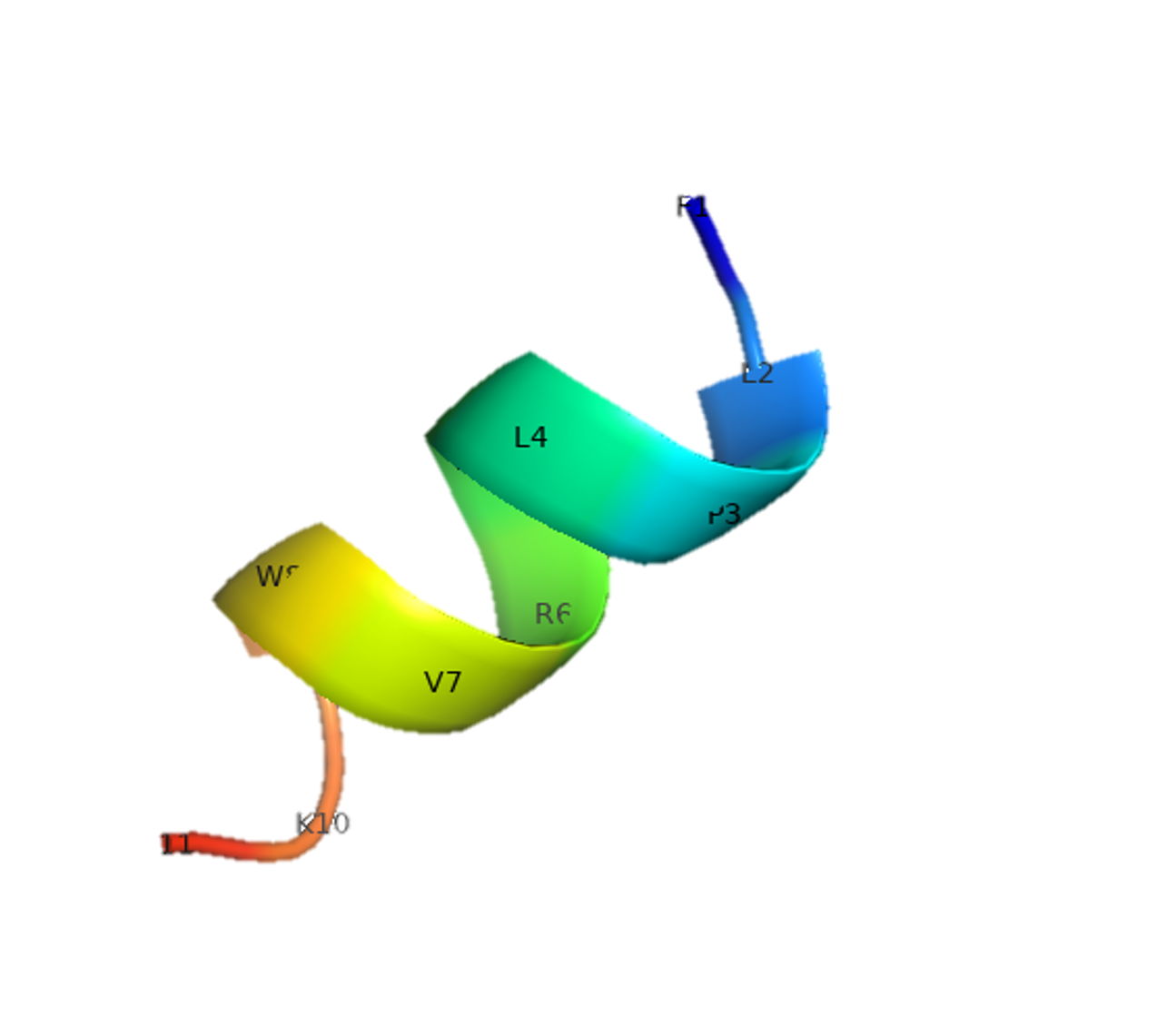}
  }
  \subfigure[Y4]{
    \label{fig:seq2}
    \includegraphics[width=0.20\linewidth]{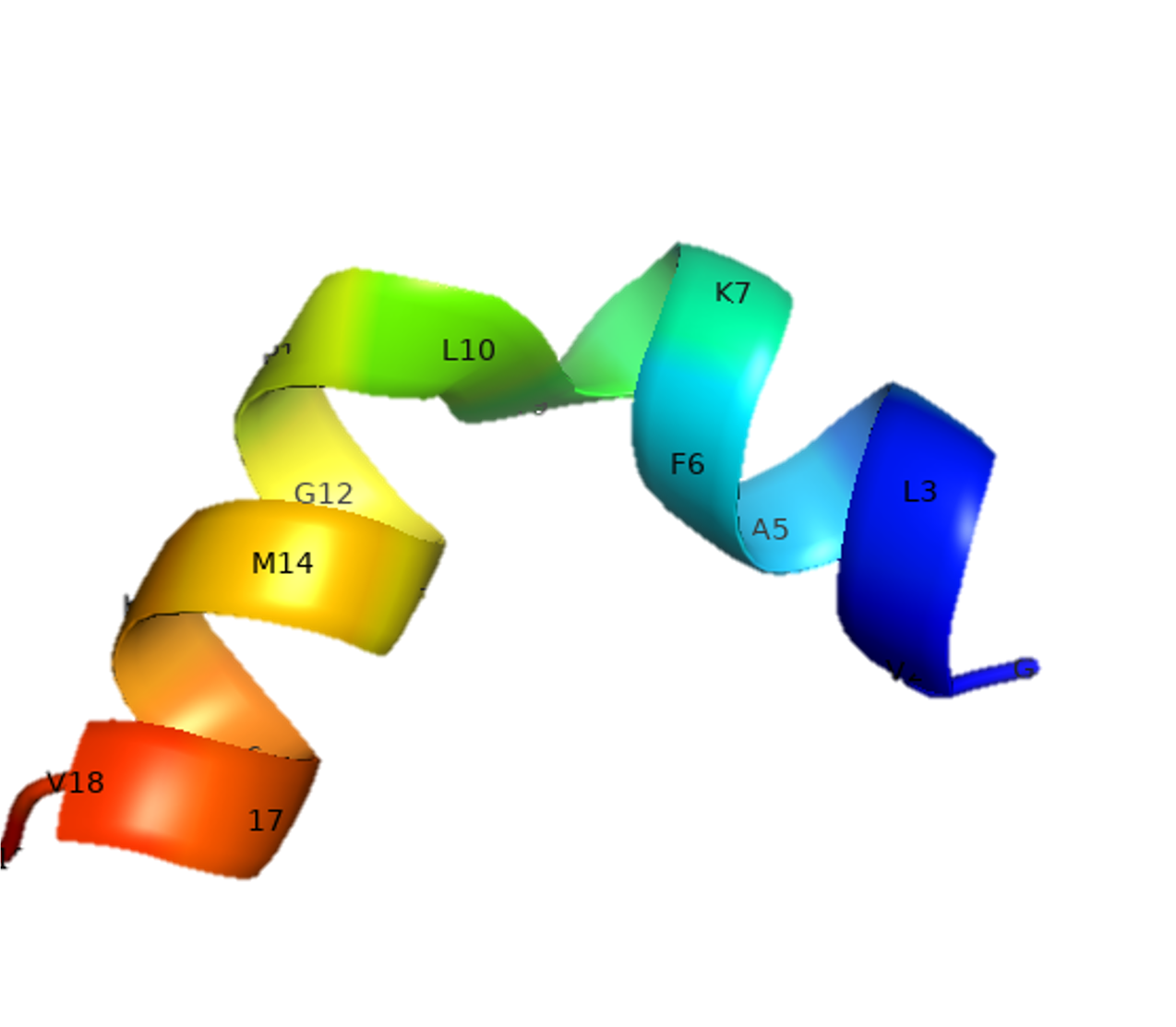}
  }
  \subfigure[Y9]{
    \label{fig:seq3}
    \includegraphics[width=0.20\linewidth]{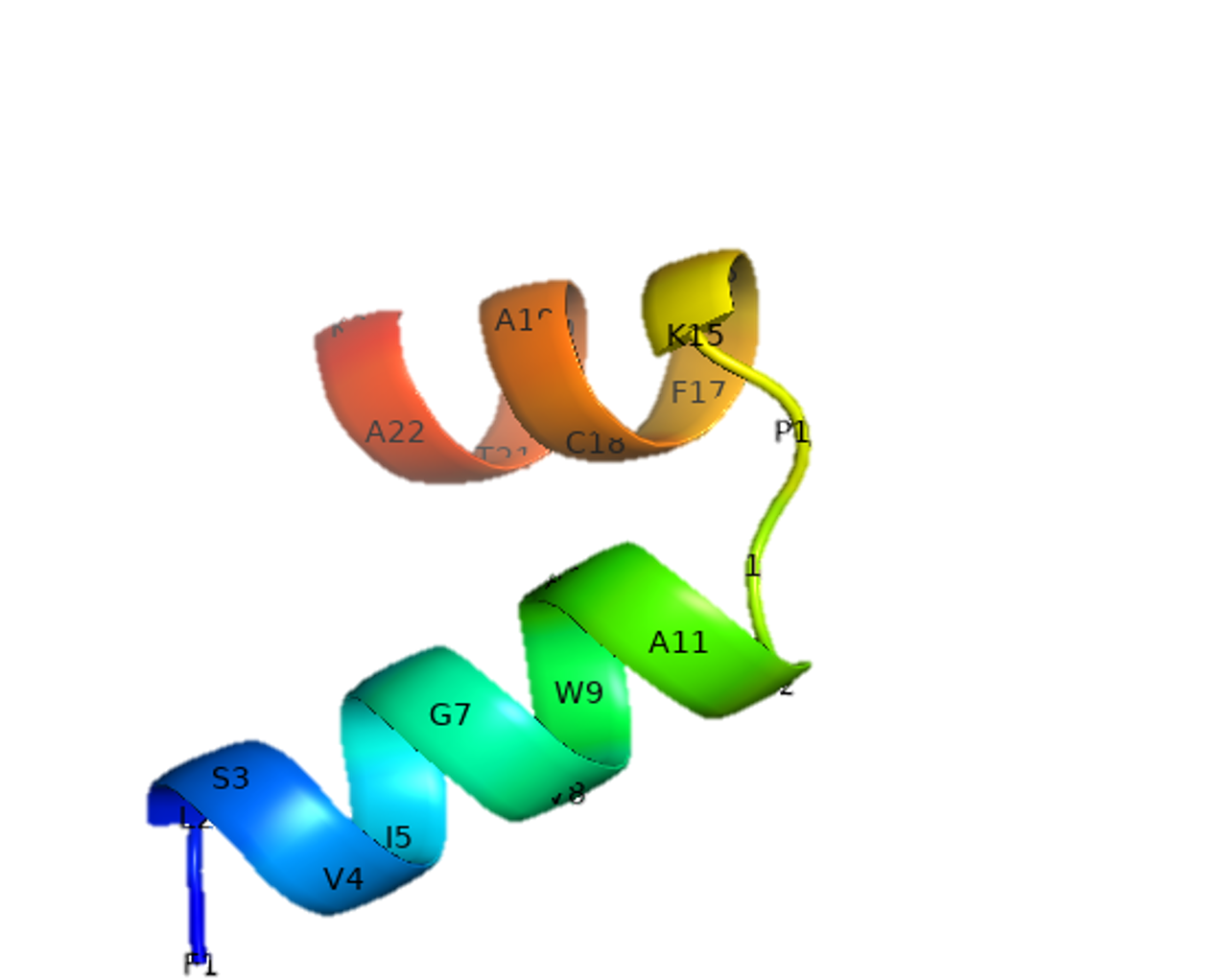}
  }
  \subfigure[Y10]{
    \label{fig:seq4}
    \includegraphics[width=0.20\linewidth]{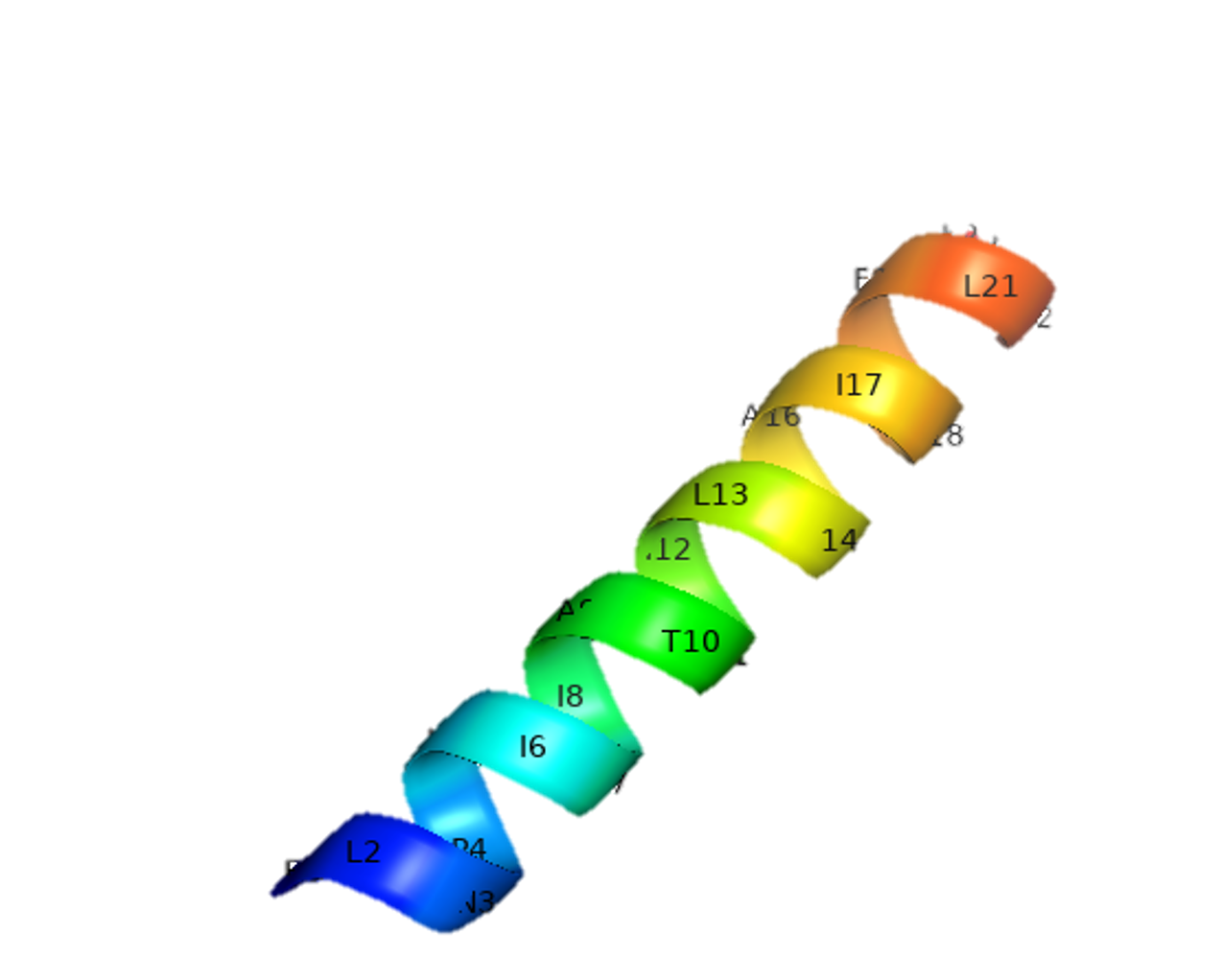}
  }
 \caption{3D structures for Y1, Y4, Y9, and Y10 in Table \ref{tab:case}.}
 \label{fig:case3d}
\end{figure}

\begin{figure*}[htbp]
    \centering
    \includegraphics[width=0.9\linewidth]{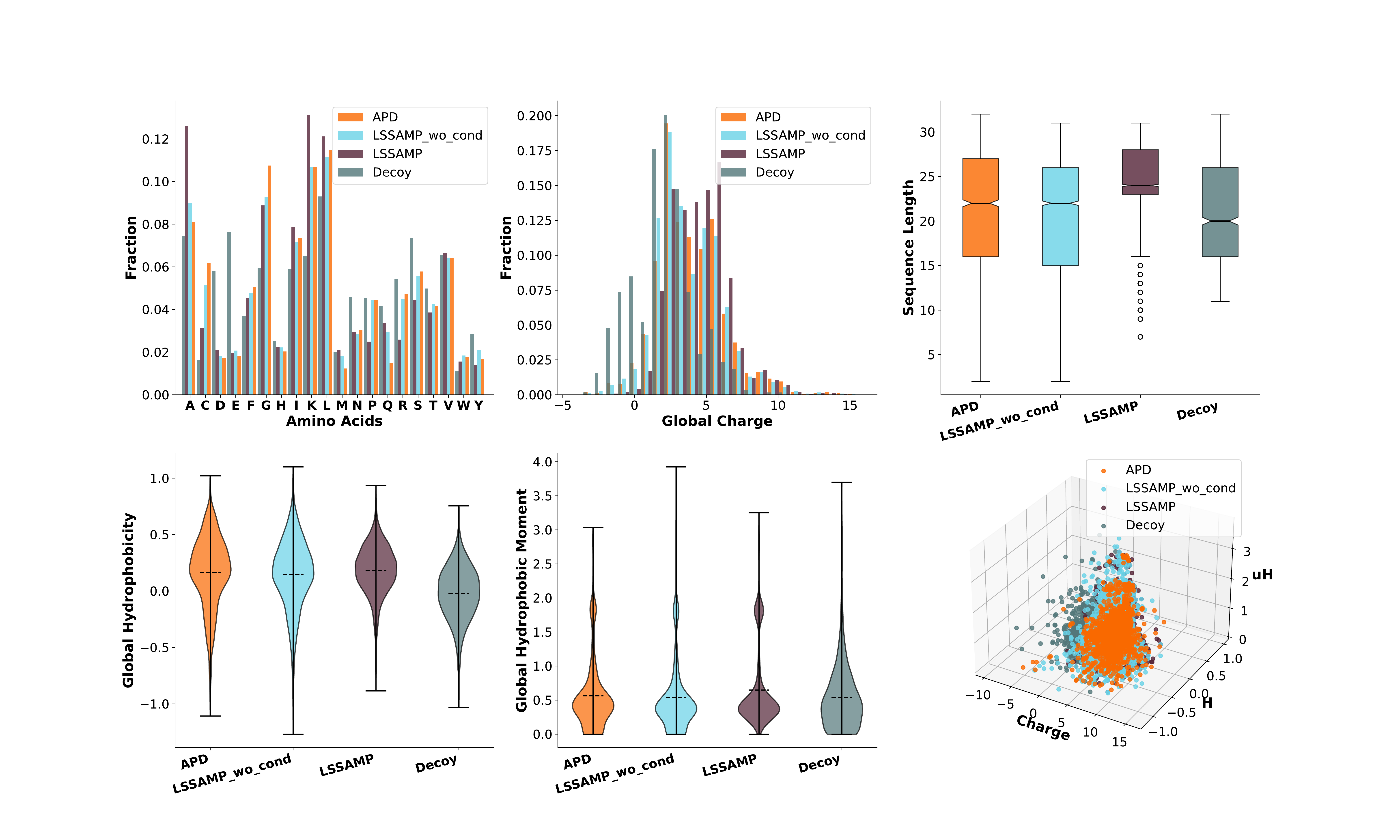}
    \caption{The distribution of amino acids, charge, sequence length, hydrophobicity, hydrophobic momentum, and a 3D visualization for three sequence attributes.}
    \label{fig:distribution}
\end{figure*}

\subsection{Analysis}
\label{sec:analysis}
\textbf{Codebook Number} 
\label{sec:book}
We explore the effect of different numbers of codebooks on generation performance. From Table \ref{tab:codebook_full}, we find that a single small codebook can hardly learn enough information to reconstruct the sequence. The PPL, Loss, and SS Acc. become better with the increase of codebook entries. However, the reconstruction accuracy achieves the best performance when the codebook is 3. This may be due to the relatively short local pattern of sequences, making the window of 8 too long for it. We do not increase the window size to $[1,2,4,8,16]$ because the maximum length of the sequence has been set to 32, making 16 too long to capture features.

\textbf{Case Study}
We show 10 peptides generated by \method in Table \ref{tab:case}. We can see that all of them have long alpha-helix in the middle and coil structures at the head or tail. We can also find that these sequences have positive charges with high hydrophobicity and hydrophobicity momentum. 
We further predict and build 3D structures of 4 generated sequences via PEPFold 3~\citep{shen2014improved} and draw the image via PyMOL~\citep{PyMOL} in Figure \ref{fig:case3d}. We can find that all these four peptides have helical structures, which is consistent with our secondary structure prediction. These helical structures make them more likely to have the antimicrobial ability. However, our model also fails for predicting a long continuous helical structure for Y4 and Y9. In fact, they have a small coil structure between the two helical structures. It indicates that our model tends to predict a long continuous secondary structure instead of several discontinuous small fragments.

\textbf{Visualization of LSSAMP Distribution}
We plot the distribution of residues, charge, sequence length, hydrophobicity, and hydrophobic momentum for APD, Decoy, and \method and \method w/o cond in Figure \ref{fig:distribution}. 
For the distribution of amino acids, the x-axis is different amino acids and the y-axis is their frequency in the generation set. We can find that without the extra condition, \method w/o cond has a similar amino acid distribution with APD. However, when we add the secondary structure conditions, it will greatly increase the frequency of A, K, and L. And we can find that those amino acids have a relatively high distribution in APD, which means that they might be responsible for the antimicrobial activity. Therefore, the addition of secondary structures may further change the direction of amino acid distribution to increase the probability of becoming AMPs.

For the global charge, we can find that compared with APD, the decoy dataset has more negative-charge sequences. And our models tend to generate sequences with positive charges. In the sequence length, we can find that the control of structure makes the length of generated sequence less diverse and tends to be longer. This is because we forced the generated sequences to have an alpha-helix structure of a length of more than 4.
In hydrophobicity and hydrophobic momentum, we can find a similar tendency where the \method w/o cond captures the distribution of APD, and the secondary structure condition makes the distribution more concentrated with a higher mean.

In the 3D visualization, we use the charge, H, and uH as the axes to see the combination of the three attributes. We can see that \method and \method w/o cond are almost overlapped by APD, which indicates that these three have a similar distribution on these attributes. Decoy sequences are out of the scope. 

To conclude, Figure \ref{fig:distribution} is a sanity check indicating our base model \method w/o cond can successfully capture the distribution of APD from various dimensions, and the extra secondary structure condition will further improve the distribution.

\subsection{Limitation}
\label{sec:limitation}
Although \method has shown its effectiveness, it is limited by several factors.
First, \method models the secondary structure instead of the 3D structure. The use of 3D structures is limited by the size of the available precise data and the difficulty of predicting 3D structures with only one input sequence.
Compared to 3D structures, secondary structures are much easier to annotate.
Benefiting from the development of AlphaFold2 in predicting 3D structures, we expect to further extend the work and incorporate the 3D information into the latent representation space and generate sequences with ideal 3D structures. 

The other limitation is that currently there is no standard evaluation metric for antimicrobial activities. 
In this paper, we follow the previous practice of using AMP classifiers and sequence properties to evaluate performance. 
However, the performance of AMP classifiers on generated peptides may be unreliable because they are trained on existing AMPs~\citep{aronica2021computational}. Furthermore, it is difficult to identify a reasonable range of sequence properties. It is because existing AMPs have different mechanisms that result in various sequence attributes. The only reliable evaluation method to check antimicrobial activity is the wet laboratory test, but it is expensive and time-consuming, which makes it impossible to perform a large number of evaluations. In the future, we hope that more reliable automatic evaluation metrics for AMPs will be proposed.

\section{Conclusion}
\label{sec:conclusion}
In this paper, we propose \method that jointly learns sequential and structural features in the same latent space and can generate peptides with ideal sequence attributes and secondary structures simultaneously. Moreover, It leverages multi-scale VQ-VAE for fine-grained control of each position. The performance evaluated using open resource AMP predictors and computational sequence attributes indicates the effectiveness of \method. It further designs two peptides with high activity against Gram-negative bacteria. 
This suggests that our generative model can effectively create an AMP library with high-quality candidates for follow-up biological experiments, which can accelerate the AMP discovery.

\section*{Acknowledgement}
\label{sec:acknowledge}
We thank Wuxi AppTec for conducting the wet laboratory experiments. The work was supported by ByteDance Research.

\newpage
\bibliographystyle{ACM-Reference-Format}
\balance
\bibliography{reference}

\newpage
\appendix
\section{Appendix}
\subsection{Attribute Distribution}
\label{sec:apd}
\begin{figure}[ht]
    \centering
    \includegraphics[width=1\linewidth]{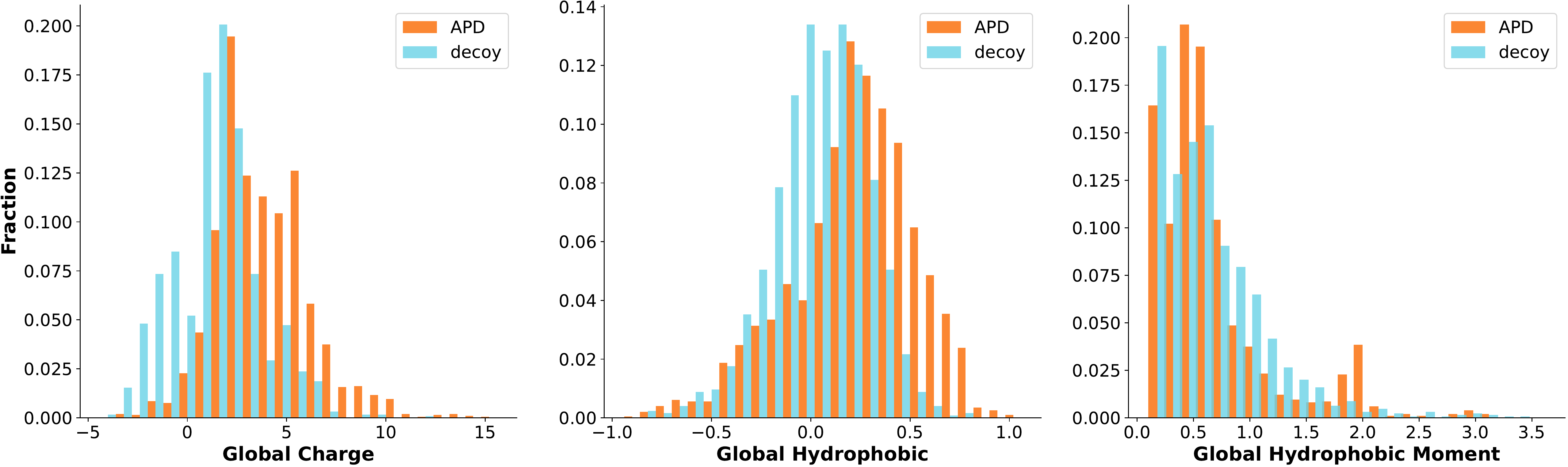}
    \caption{The histogram of charge, hydrophobicity and hydrophobic moment on APD and decoy dataset. }
    \label{fig:AMP}
\end{figure}

To determine the effective threshold of charge, hydrophobicity, and hydrophobic moment of AMP, we analyze the sequence distribution in APD and decoy in Figure \ref{fig:AMP}. For charge, we follow the rule summarized by experts and choose sequences whose net charge is +2 to +10. For the remaining two characters, we draw a histogram and compare the proportion in each box. If the proportion of APD is larger than that in the decoy, we add bin to the acceptance range of the evaluation metric. The final ranges are $C \in [2,10]$, $ H \in [0.25, \infty]$, and $uH \in [0.5,0.75] \cup [1.75, \infty]$.

\begin{figure}
    \centering
    \includegraphics[width=1\linewidth]{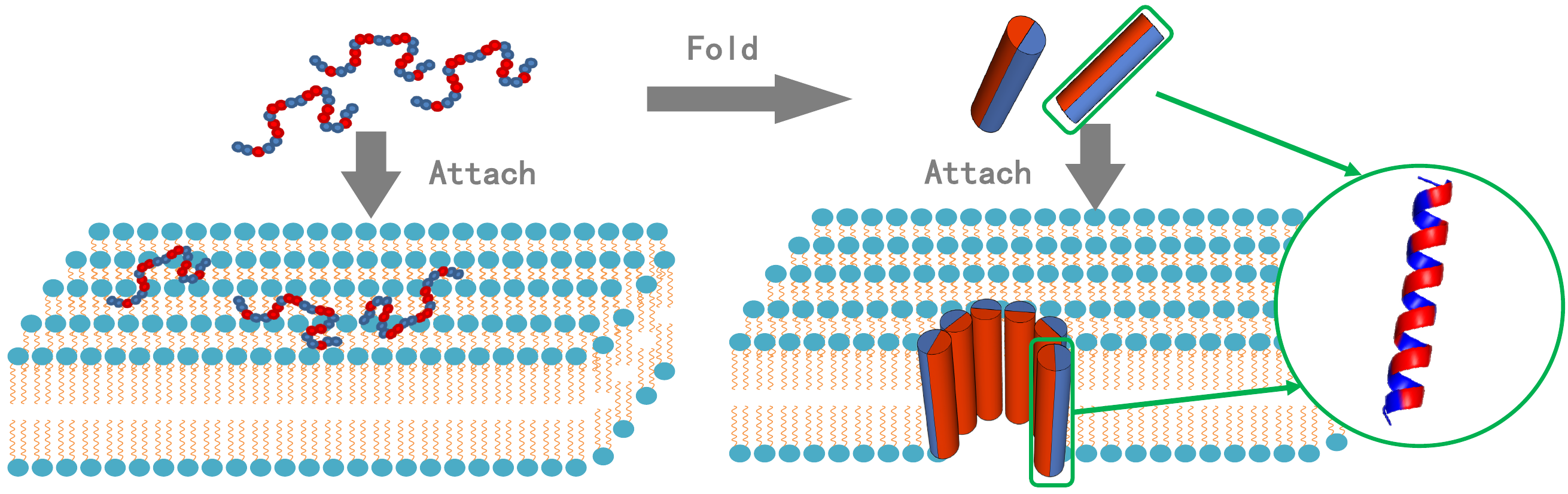}
    \caption{An example of the antimicrobial mechanism. The blue indicates the hydrophobic amino acids, and the red ones are hydrophilic. On the left, although the peptides with reasonable amino acids have attached to the bacterial membrane, they still can not insert into it. However, by folding into the helix structure, as shown on the right, the peptides maintain a stable hole that breaks the membrane of the bacterium.}
    \label{fig:mechanism}
\end{figure}

\begin{table*}[tb]
\setlength{\tabcolsep}{3pt}
  \centering
    \begin{tabular}{lccccc}
    \toprule
      & \multicolumn{1}{c}{\textbf{Uniq}} & \multicolumn{1}{c}{\textbf{C}} & \multicolumn{1}{c}{\textbf{H }} & \multicolumn{1}{c}{\textbf{uH}} & \multicolumn{1}{c}{\textbf{Combination}} \\
    \midrule
    \textbf{Random $p = 0.1$} & 2055 & 7.38\% $\pm$ 11.01\% & 37.93\% $\pm$ 0.44\% & 4.61\%  $\pm$ 0.31\% & 4.34\%  $\pm$ 0.41\% \\
    \textbf{Random $p = 0.2$} & 1831 & 6.87\% $\pm$ 0.31\% & 9.52\% $\pm$ 0.31\% & 1.91\% $\pm$ 1.06\% & 2.19\% $\pm$ 0.66\% \\
    \textbf{VAE}~\citep{dean2020variational} & 475 & 18.45\% $\pm$ 2.92\% & 2.68\% $\pm$ 3.28\% & -2.78\% $\pm$ 1.64\% & 0.29\% $\pm$ 0.74\% \\
    \textbf{AMP-GAN}~\citep{van2021ampgan} & 1966 & 2.79\% $\pm$ 0.50\% & 2.16\% $\pm$ 0.34\% & -2.29\% $\pm$ 0.53\% & 0.17\% $\pm$ 0.35\% \\
    \textbf{PepCVAE}~\citep{das2018pepcvae} & 208 & 3.87\% $\pm$ 1.58\% & -1.93\% $\pm$ 1.61\% & 1.01\% $\pm$ 2.80\% & 3.93\% $\pm$ 1.82\% \\
    \textbf{MLPeptide}~\citep{capecchi2021machine} & 2106 & -2.48\% $\pm$ 0.39\% & 2.01\% $\pm$ 0.57\% & 9.24\% $\pm$ 1.22\% & 1.12\% $\pm$ 0.38\% \\
    \textbf{\method} & 4876 & 0.30\% $\pm$ 0.37\% & 3.96\% $\pm$ 0.64\% & 7.53\% $\pm$ 0.41\% &  1.87\% $\pm$ 0.07\% \\
    \bottomrule
    \end{tabular}%
  \caption{The delta ratio of sequence properties filtered by secondary structures. \textbf{Uniq} is the uniq peptide number among 5000 generated sequences. \textbf{C}, \textbf{H}, \textbf{uH} correspond to charge, hydrophobicity, hydrophobic moment. \textbf{Combination} is the percentage satisfying three ranges at the same time.}
  \label{tab:filtered_full}%
\end{table*}%

\subsection{Secondary Structure Filter}
Similar to proteins, the biological functions of AMPs are determined by their amino acid sequences and folded structures~\citep{boman2003antibacterial}. If the peptide can not fold into an appropriate structure, it is still difficult to take effect. For example, by forming a helical structure, the peptide can gather hydrophobic amino acids on one side and hydrophilic amino acids on the other.
This amphiphilic structure helps the peptide insert into the membrane and maintain a stable hole with other molecules in the membrane, as shown in Figure \ref{fig:mechanism}. Without it, the peptide can hardly penetrate the membrane and attach to the surface. 

\textit{But does controlling secondary structure also affect sequence attributes?} To answer this question, we control the secondary structure of the generated peptides to $\alpha$-helix for our baseline. The performance gaps are shown in Table \ref{tab:filtered_full}. From Table \ref{tab:filtered_full}, we can find that most of the results are improved by limiting sequences to the $\alpha$-helix structures. It shows that by controlling the structure, the sequence attributes can be improved, which verifies the importance of introducing secondary structures to the controlled generation process. However, the sequence size has decreased significantly, indicating that this generate-then-filter pipeline is inefficient.

\begin{figure}[ht]
    \centering
    \includegraphics[width=0.7\linewidth]{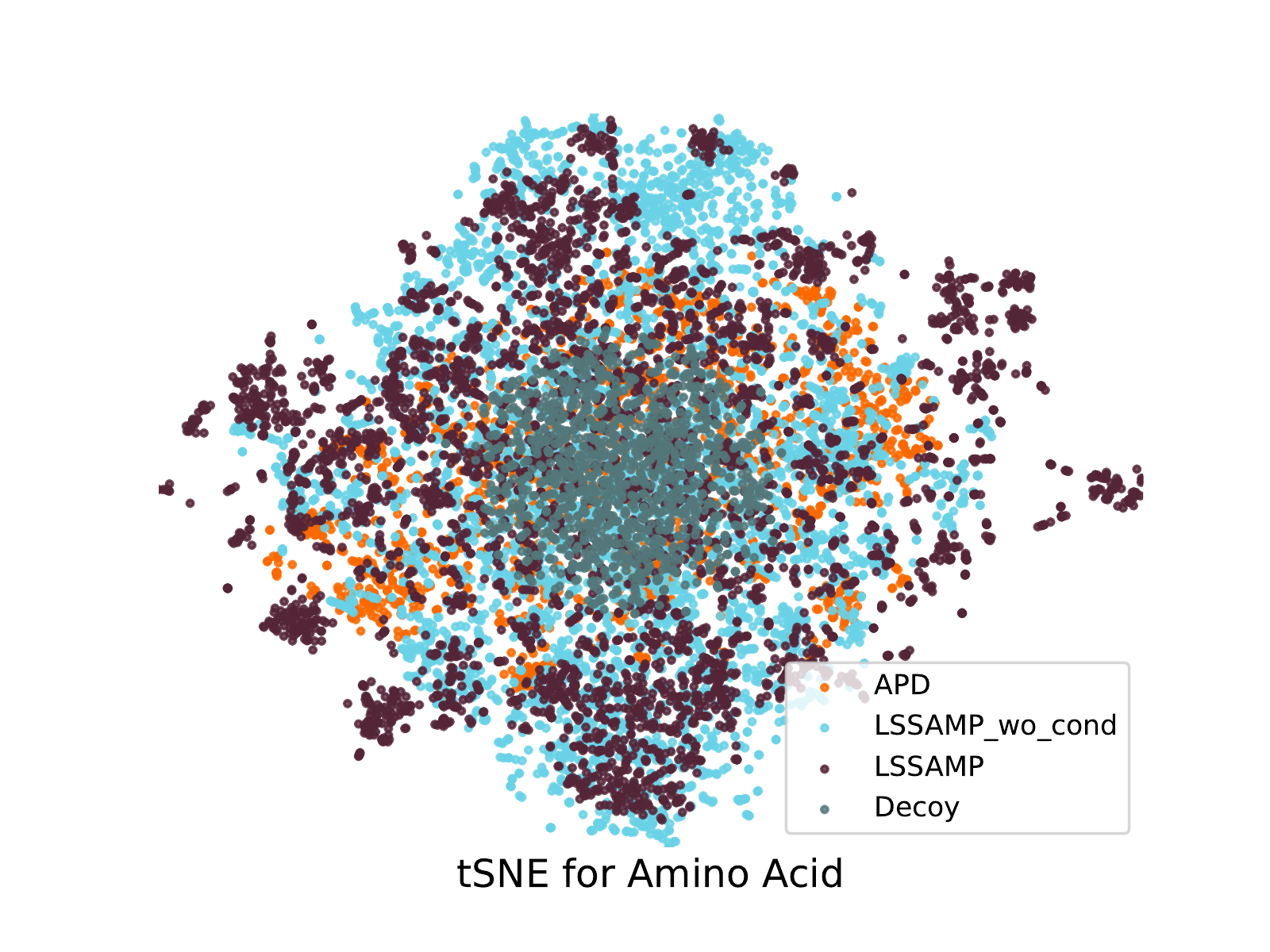}
    \caption{The tSNE plot for the distribution of residue in each sequence on four datasets.}
    \label{fig:aafreq}
\end{figure}

\paragraph{Visualization of Residue Distribution}
To illustrate the distribution of residues in the generated peptides, we plot tSNE, shown in Figure \ref{fig:aafreq}. We transform the vector with each dimension representing the probability of a certain residue to represent the peptide. Then we use tSNE to convert the high-dimensional vector to 2D and visualize them. We find that there is a large overlap between \method w/o condition and APD, which indicates that our model has captured the global distribution of APD instead of collapsing to a local mode. Furthermore, \method covers APD and has some outliers. The results show that with the secondary structure condition, our model can not only learn the existing AMP distribution but also explore more possible spaces.

\begin{figure}
    \centering
    \includegraphics[width=0.75\linewidth]{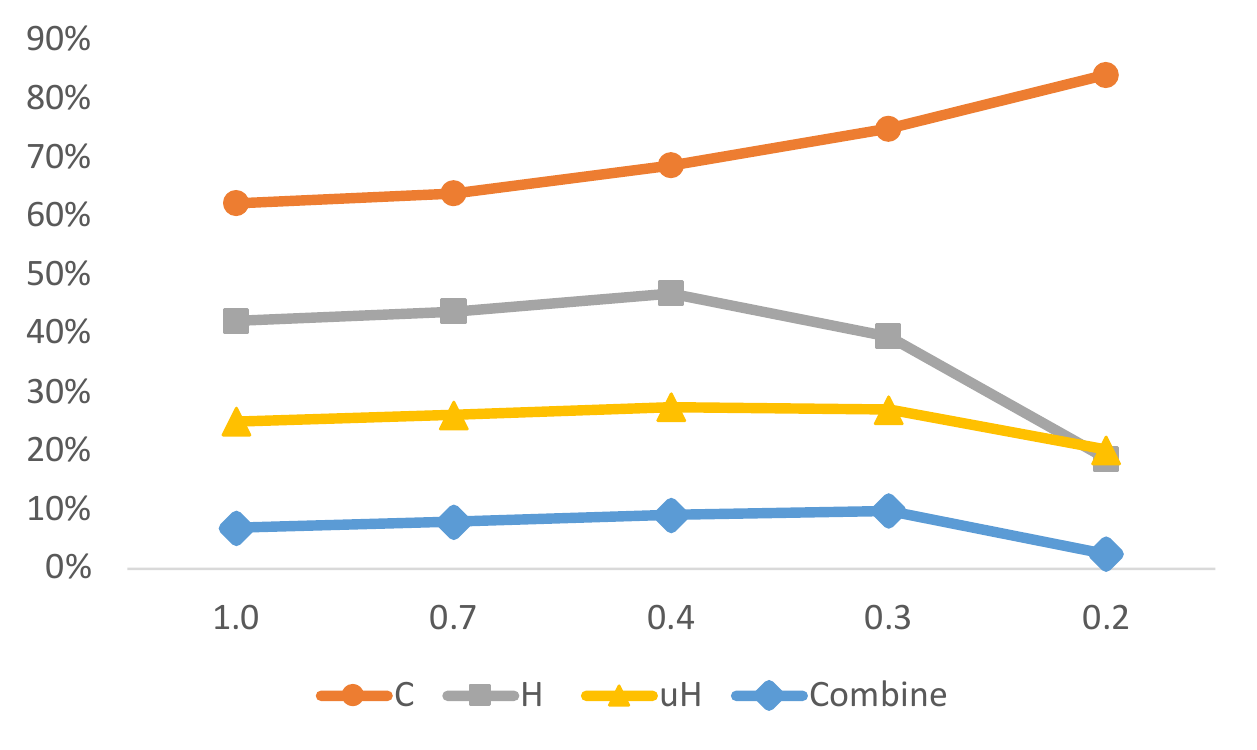}
    \caption{The sequence attributes of peptides with different percentages of the coil structure. The x-axis is the maximum percentage and the y-axis is the percentage of peptides that meet the attribute range.}
    \label{fig:condition}
\end{figure}

\paragraph{Structure Condition} 
As described above, controlling the secondary structure can affect the attributes of generated peptides. Thus we limit the percentage of the coil structure with different ratios and calculate the sequence attributes of generated peptides. The results are shown in Figure \ref{fig:condition}. The x-axis is the maximum percentage of the coil structures allowed during the generation. We can find that with the decrease in the length of coil structures, the percentage of positive peptides keeps growing. 
However, for hydrophobicity and hydrophobic momentum, the percentage drop after 0.3. 
Therefore, we limit the length of the coil structure to 30\% in our main experiments.

\subsection{Implementation}
\label{sec:implementation}
\textbf{Implementation details} There are three main modules for \method. The encoder and decoder are based on a 2-layer Transformer~\citep{vaswani2017attention} with $d_{model} = 128$, $head=8$. The size of FFN projection is $d_{ffn}=512$ and all drouput rate are $0.1$. For the classifier, we use the same CNN block as \citet{billings2019prospr} with 32 input channels and a dilation scale of $[1,2,4,8,10]$. For multi-scale codebooks, we first apply CNN to extract features. We set $n=4$ and kernel width ranging in $[1,2,4,8]$. The features will be padded to the same length as the input sequence. Then, we use 4 codebooks with $K=8$ and $d=128$. The reconstruction and prediction share the same codebooks, which means $N_{r}=N_{s}=4$. The commit coefficient is set to $\beta=0.05$. 

\textbf{Reproduction}
We run the model several times and calculate the mean and variance of the main experimental results and analysis. We use PyTorch to implement our model and train it on a single Tesla-V100-32GB. We optimize the parameter with Adam Optimizer~\citep{kingma2015adam}. During pre-training for sequence construction on $D_{r}$, we set the maximum token in a batch $bz$ as 30,000, learning rate $lr$ as 0.01 with 8,000 warm-up steps, and decoy weight for EMA as $\lambda=0.8$. For secondary structure prediction on $D_{s}$, the max length is limited to 32, $bz=10,000$, $lr=0.003$, $\lambda=0.95$, and the prediction loss coefficient $\gamma=1$. Finally, we transfer to $D_{amp}$ with the same hyperparameters except the $lr=0.001$.

\subsection{More Case Study}
\begin{figure}[ht]
  \centering
  \subfigure[Y2]{
    \label{fig:seq5}
    \includegraphics[width=0.40\linewidth]{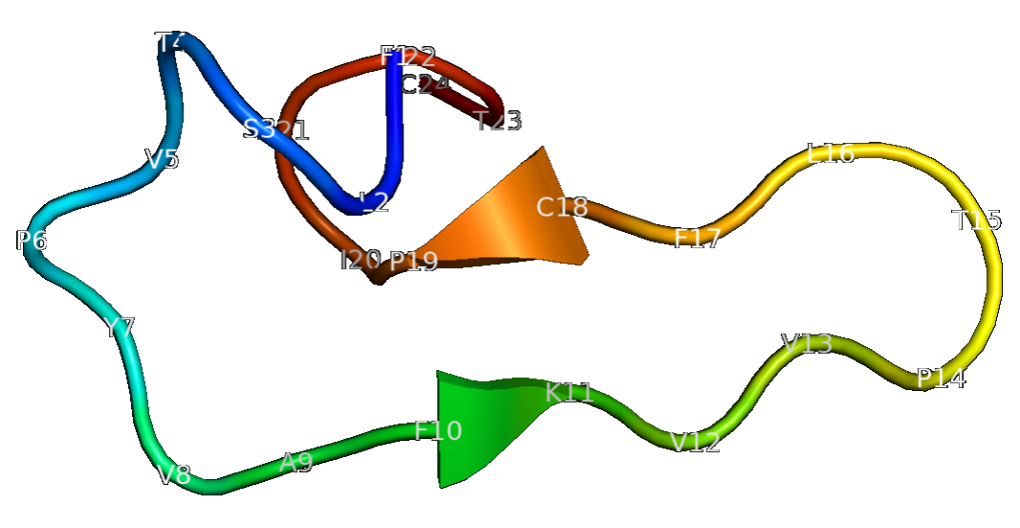}
  }
  \subfigure[Y6]{
    \label{fig:seq6}
    \includegraphics[width=0.40\linewidth]{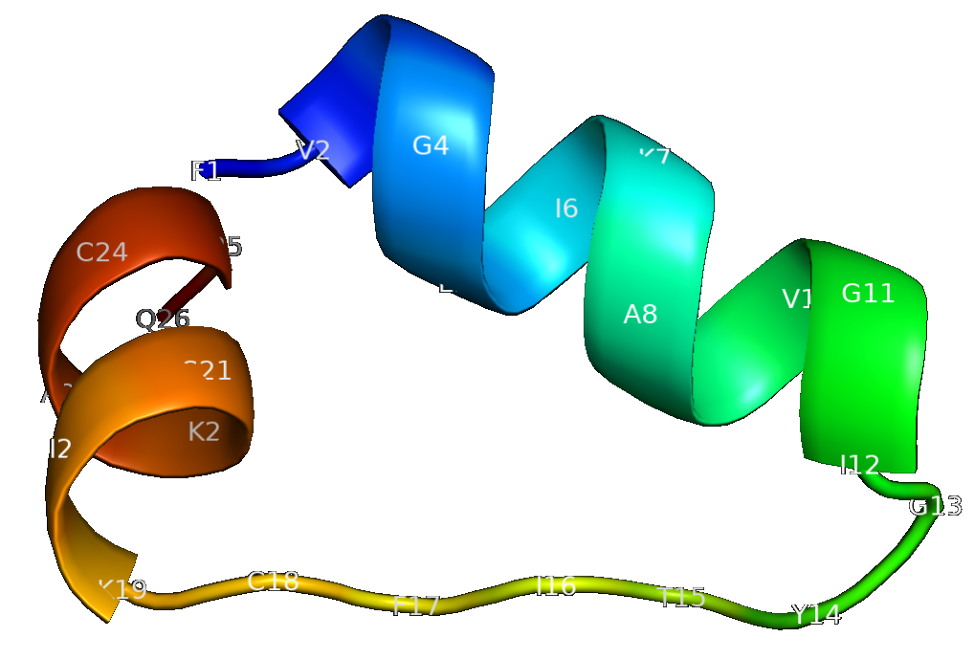}
  }
 \caption{3D structures for Y1 and Y6 in Table \ref{tab:case}.}
 \label{fig:case3d_2}
\end{figure}

\paragraph{Disulfide Bonds} Studies have shown that disulfide bonds also contribute to the native-folded AMP stability and affect the activity of AMPs by influencing their folding stability~\citep{nehls2020influence,deplazes2020unusual}. Therefore, we also manually check the disulfide bonds in Table \ref{tab:case}. For disulfide bonds, there should be at least two Cys residues, so Y2, Y6, Y9, and Y10 are likely to form disulfide bonds. Therefore, we also model 3D structure of Y2 and Y6 in Figure \ref{fig:case3d_2}. The Y9 and Y10 are shown in Figure \ref{fig:case3d}. For Y6, Y9, and Y10, at least one of the Cys residues lies on the alpha-helix and does not form the disulfide bond. For Y2, the Cys residues don’t form any stable structure. This is no surprise because we control the generated peptide to form the helix structure. Meanwhile, our model fails in predicting the coil and beta structure in Y2, which illustrates the limitation of our secondary structure modeling.

\end{document}